\documentclass[aps,prm,superscriptaddress,twocolumn,showpacs]{revtex4-1}
\usepackage{graphicx}
\usepackage{amsmath}
\usepackage{epsfig}
\usepackage{multirow}
\usepackage{array}
\usepackage{inputenc}
\usepackage{color}
\usepackage{tabularx}
\usepackage{textcomp}
\usepackage{ulem}
\usepackage[colorlinks=true,linkcolor=black, citecolor=blue, urlcolor=blue, unicode=true]{hyperref}

\begin{document}

\title{Lifetime shortened acoustic phonons and static order at the Brillouin zone boundary in the organic-inorganic perovskite CH$_3$NH$_3$PbCl$_3$}

\author{M. Songvilay}
\affiliation{School of Physics and Astronomy and Centre for Science at Extreme Conditions, University of Edinburgh, Edinburgh EH9 3FD, UK}
\author{M. Bari}
\affiliation{Department of Chemistry and 4D labs, Simon Fraser University, Burnaby, British Columbia, Canada V5A1S6}
\author{Z.-G. Ye}
\affiliation{Department of Chemistry and 4D labs, Simon Fraser University, Burnaby, British Columbia, Canada V5A1S6}
\author{Guangyong Xu}
\affiliation{NIST Center for Neutron Research, National Institute of Standards and Technology, 100 Bureau Drive, Gaithersburg, Maryland, 20899, USA}
\author{P. M. Gehring}
\affiliation{NIST Center for Neutron Research, National Institute of Standards and Technology, 100 Bureau Drive, Gaithersburg, Maryland, 20899, USA}
\author{W. D. Ratcliff}
\affiliation{NIST Center for Neutron Research, National Institute of Standards and Technology, 100 Bureau Drive, Gaithersburg, Maryland, 20899, USA}
\author{K. Schmalzl}
\affiliation{Forschungszentrum J\"ulich GmbH, J\"ulich Centre for Neutron Science at ILL, 71 avenue des Martyrs, 38000 Grenoble, France}
\author{F. Bourdarot}
\affiliation{Universit\'e Grenoble Alpes, CEA, INAC, MEM/MDN, F-38000 Grenoble, France}
\author{B. Roessli}
\affiliation{Laboratory for Neutron Scattering and Imaging (LNS), Paul Scherrer Institut (PSI), 5232 Villigen PSI, Switzerland}
\author{C. Stock}
\affiliation{School of Physics and Astronomy and Centre for Science at Extreme Conditions, University of Edinburgh, Edinburgh EH9 3FD, UK}
\date{\today}

\begin{abstract}

Lead halide hybrid perovskites consist of an inorganic framework hosting a molecular cation located in the interstitial space. These compounds have been extensively studied as they have been identified as promising materials for photovoltaic applications with the interaction between the molecular cation and the inorganic framework implicated as influential for the electronic properties. CH$_3$NH$_3$PbCl$_3$ undergoes two structural transitions from a high temperature cubic unit cell to a tetragonal phase at 177 K and then a subsequent orthorhombic transition at 170 K.  We have measured the low-frequency lattice dynamics using neutron spectroscopy and observe an energy broadening in the acoustic phonon linewidth towards the high-symmetry point $\textbf{Q}_X=(2,\frac{1}{2}, 0)$ when approaching the transitions.  Concomitant with these zone boundary anomalies is a hardening of the entire acoustic phonon branch measured in the ${\bf{q}}\rightarrow 0$ limit near the (2, 0, 0) Bragg position with decreasing temperature.   Measurements of the elastic scattering at the Brillouin zone edges $\textbf{Q}_X=(2,\frac{1}{2}, 0)$, $\textbf{Q}_M=(\frac{3}{2}, \frac{1}{2}, 0)$, and $\textbf{Q}_R=(\frac{3}{2},\frac{3}{2}, \frac{5}{2})$ show Bragg peaks appearing below these structural transitions.  Based on selection rules of neutron scattering, we suggest that the higher 177 K transition is displacive with a distortion of the local octahedral environment and the lower transition is a rigid tilt transition of the octahedra.  We do not observe any critical broadening in energy or momentum, beyond resolution, of these peaks near the transitions.  We compare these results to the critical properties reported near the structural transitions in other perovskites and particularly CsPbCl$_{3}$ (Ref. \onlinecite{Fujii1974}).  We suggest that the simultaneous onset of static resolution-limited Bragg peaks at the zone boundaries and the changes in acoustic phonon energies near the zone center is evidence of a coupling between the inorganic framework and the molecular cation.  The results also highlight the importance of displacive transitions in organic-inorganic hybrid perovskites.  

\end{abstract}


\maketitle

\section{Introduction}

\begin{figure}
 \includegraphics[scale=0.33]{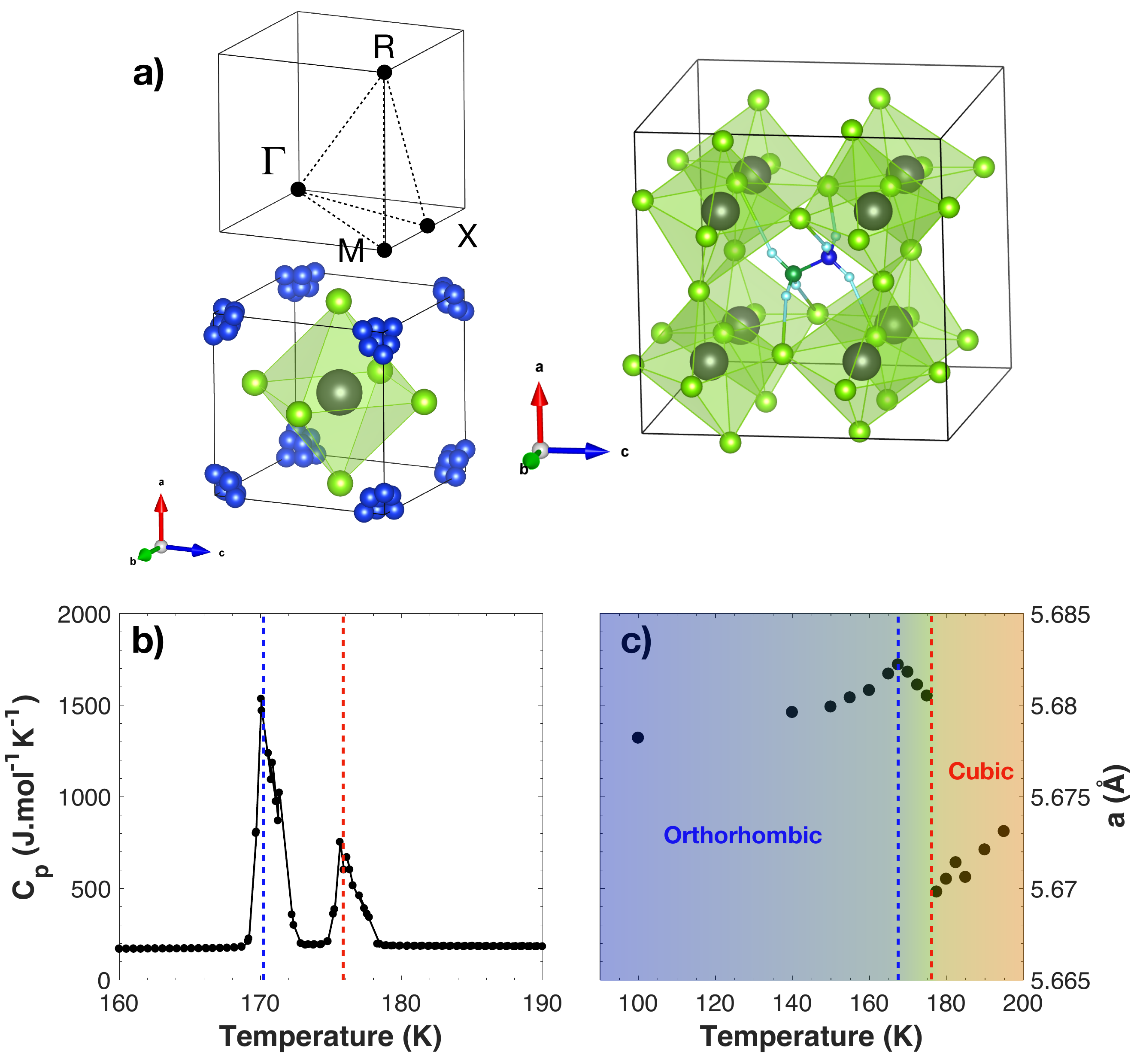}
 \caption{ \label{fig:structure} $(a)$ left: Crystallographic structure of MAPbCl$_3$ in the cubic phase. The Pb site is represented by the grey sphere, the Cl ions are shown in green, and the methylammonium molecules are represented in blue in their disordered state. The top figure shows the high symmetry points in the first Brillouin zone of a primitive cubic lattice. $X$, $M$ and $R$ represent $\textbf{Q}=(\frac{1}{2},0, 0), (\frac{1}{2},\frac{1}{2}, 0)$, and $(\frac{1}{2},\frac{1}{2}, \frac{1}{2})$, respectively. right: Crystallographic structure of MAPbCl$_3$ in the orthorhombic phase. In this ordered phase, the H, N and C atoms forming methylamonium molecules are represented with light blue, dark blue and dark green spheres, respectively. $(b)$ Heat capacity as a function of temperature for MAPbCl$_3$.  $(c)$ Temperature dependence of the cell parameter $a$, measured on BT4 (NIST). The dashed lines indicate the two structural transitions in MAPbCl$_3$. }
\end{figure}

Organic-inorganic hybrid halide perovskites have been the subject of extensive theoretical and experimental work as promising materials for photovoltaic and optoelectronic devices \cite{Saparov2016,Brenner2016,Charles2017,Ghosh2017}. These systems can be described with the well-known ABX$_{3}$ perovskite structure where the A site is occupied by a molecular cation (most commonly methylammonium CH$_{3}$NH$_{3}$ = MA or formamidinum HC(NH$_{2}$)$_{2}$ = FA),  and B is a metal (Pb or Sn) located in an octahedral environment provided by the X-site (Cl, Br or I), thus forming an inorganic framework. It should be emphasized that the molecular cation is strongly coupled to the inorganic framework through hydrogen bonding as shown in Fig. \ref{fig:structure} $(a)$ \cite{Lee2015}.  In particular, MAPbI$_{3}$ has attracted a lot of attention as solar cells made of this material were found to reach a high power conversion efficiency exceeding 20\% \cite{Saparov2016,Ren2016,Weller2015}. Thus, fundamental studies of the structural dynamics are key to understanding the electronic properties. We address here the case of the chlorine compound MAPbCl$_{3}$.

Along with its inorganic counterparts ~\cite{Fujii1974, Hirotsu1974, Hua1991}, MAPbCl$_{3}$ undergoes several structural transitions on cooling\cite{Egger2016} with a cubic to tetragonal distortion at 177~K and then to an orthorhombic phase below 170~K. These transition temperatures were confirmed in our single crystal sample by heat capacity measurements presented in Fig. \ref{fig:structure} $(b)$. In the high temperature cubic phase, the site symmetry of the A cation exceeds the molecular symmetry, which forces the MA cations to be disordered. Therefore, in addition to the framework distortion, the organic cation undergoes an order-disorder transition associated with molecular dynamics. In MAPbBr$_{3}$, neutron quasielastic scattering studies highlight the onset of the molecular dynamics in two steps: a transition occurring at the tetragonal-orthorhombic transition (T = 150 K) from a rotation of the full molecule \cite{Brown2017,Swainson2015} to a state of rotations of the CH$_3$ and NH$_3$ groups around the C-N axis, and a second transition where the molecule reaches a static order upon cooling. In addition, I. P. Swainson \textit{et al.} \cite{Swainson2015} showed that the first high-temperature transition is concomitant with a phonon soft mode, associated with the PbBr$_3$ framework, thereby demonstrating the coupling between the organic cation and the inorganic framework.   However, we emphasize that these measurements were performed on powder samples and therefore the exact reciprocal lattice point associated with this softening could not be identified.

The crystal structure of MAPbCl$_{3}$ was first characterized by Poglitsch and Weber \cite{Potglisch1987} who proposed the cubic phase to be $Pm3m$ ($a$ = 5.675 \AA), the tetragonal phase to be $P4/mmm$ ($a$ = 5.656 \AA, $c$ = 5.630 \AA), and the orthorhombic phase to be $P222_1$ ($a$ = 5.673 \AA, $b$ = 5.628 \AA, $c$ = 11.182 \AA).  The structure is shown in Fig. \ref{fig:structure} $(a)$.  L. Chi \textit{et al.} \cite{Chi2005} later suggested that the space group for the orthorhombic phase is $Pnma$ ($a$ = 11.1747 \AA, $b$ = 11.3552 \AA, $c$ = 11.2820 \AA), based on powder diffraction measurements. A single crystal diffraction study of the tetragonal phase was also carried out by Kawamura and Mashiyama \cite{Kawamura1999, Swainson2005} who observed both superlattice and incommensurate reflections in this intermediate phase, as was also reported for MAPbBr$_{3}$ \cite{Guo2017}. 

Many studies have been devoted to the molecular dynamics in MAPbX$_3$ \cite{Ren2016,Leguy2015,Brown2017,Weller2015,Even2016}, but investigating the coupling between the molecular dynamics and the inorganic framework is key to understanding the mechanisms involved in their photovoltaic properties. Indeed, although the MA cation does not influence directly the electronic band structure, it was suggested that the orientational dynamics of the molecule and its coupling to the inorganic framework through acoustic phonon modes allow longer carrier lifetime and mobility, that exceeds those of currently used crystalline semi-conductors \cite{Brivio2015,Zhu2016, Swainson2015, Brown2017}. 
In this work, we will show the presence of strong acoustic dampening and also changes in energy near the structural transition.  Based on this we suggest a coupling between framework and rotational dynamics in MAPbCl$_{3}$. 

This manuscript is divided in five sections including this introduction. The experimental details are presented in section two, and sections three and four present the neutron inelastic measurements of acoustic phonons at 300 K mapping out the dispersion and temperature dependence, respectively. These results are then discussed in the last section and compared to the purely inorganic CsPbCl$_{3}$ analogue and structural phase transitions in other perovskites.

\section{Experimental details}

Neutron inelastic spectroscopy was performed on the thermal triple-axis spectrometer IN22 (ILL, Grenoble) with constant $k_f$ = 2.662 \AA$^{-1}$, using a PG filter between the sample and the analyzer to remove higher order contamination. Further measurements at several temperatures were performed on the thermal triple-axis instrument BT4 (NIST, Gaithersburg) with constant $k_f$~=~2.662~\AA$^{-1}$ using PG filters between the monochromator and the sample, and between the sample and the analyzer. Measurements were also carried out on the cold triple-axis spectrometer SPINS (NIST, Gaithersburg) with $k_f$ = 1.55 \AA$^{-1}$ using a Be filter between the sample and the analyzer. Finally, the temperature dependence of the elastic line at $\textbf{Q}_X=(2,\frac{1}{2}, 0)$ and $\textbf{Q}_M=(\frac{3}{2},\frac{1}{2}, 0)$ was measured on the cold triple-axis TASP (PSI, Villigen) with $k_f$ = 1.55 \AA$^{-1}$  using a Be filter between the sample and the analyzer. 

All phonon measurements were performed in the (H~K~0) scattering plane. As super-lattice reflections are expected at the $X$, $M$ and $R$ zone boundaries, measurements were performed towards specific directions where the structure factor was non-zero, as described in \cite{Fujii1974}.  Throughout the paper we use the zone boundary notation taken with respect to the cubic unit cell shown in Fig. \ref{fig:structure} $(a)$.  Diffraction measurements investigating the three zone boundaries were done both in (H~K~0) and (H~H~L) plane. With the 0.1 g sample aligned in the (H~K~0) plane, the phonon dispersions of the TA$_1$ and TA$_2$ modes could be extracted. The TA$_1$ mode corresponds to acoustic phonons propagating along the [1~0~0] direction with a polarization along [0~1~0] or [0~0~1]. Using the equations of motion outlined in Ref. \onlinecite{Dove1993}, the velocity of this phonon can be related to the $C_{44}$ elastic constant. The TA$_2$ phonon mode propagates along [1~1~0] with a polarization along [1~$\overline{1}$~0], and the slope of the dispersion depends on ($C_{11}-C_{12}$)/2. Given that the neutron cross section for phonon scattering scales as $(\vec{Q}\cdot \vec{\xi})^2$, where $\vec{\xi}$ is the phonon eigenvector, transverse scans near $\vec{Q}=(2~0~0)Ê\pm (0~q~0)$ afford a measurement of the TA$_1$ phonon while scans near $\vec{Q} = (2~2~0) \pm (-q~q~0)$ give the TA$_2$ mode. 

Heat capacity measurements were performed using a Physical Property Measurement System (PPMS, Quantum Design) in a temperature range between 140~K and 200~K. A relaxation method with a 2$\tau$ fitting procedure was used. 

\section{Phonon dispersion at T = 300 K}

\begin{figure}
 \includegraphics[scale=0.36]{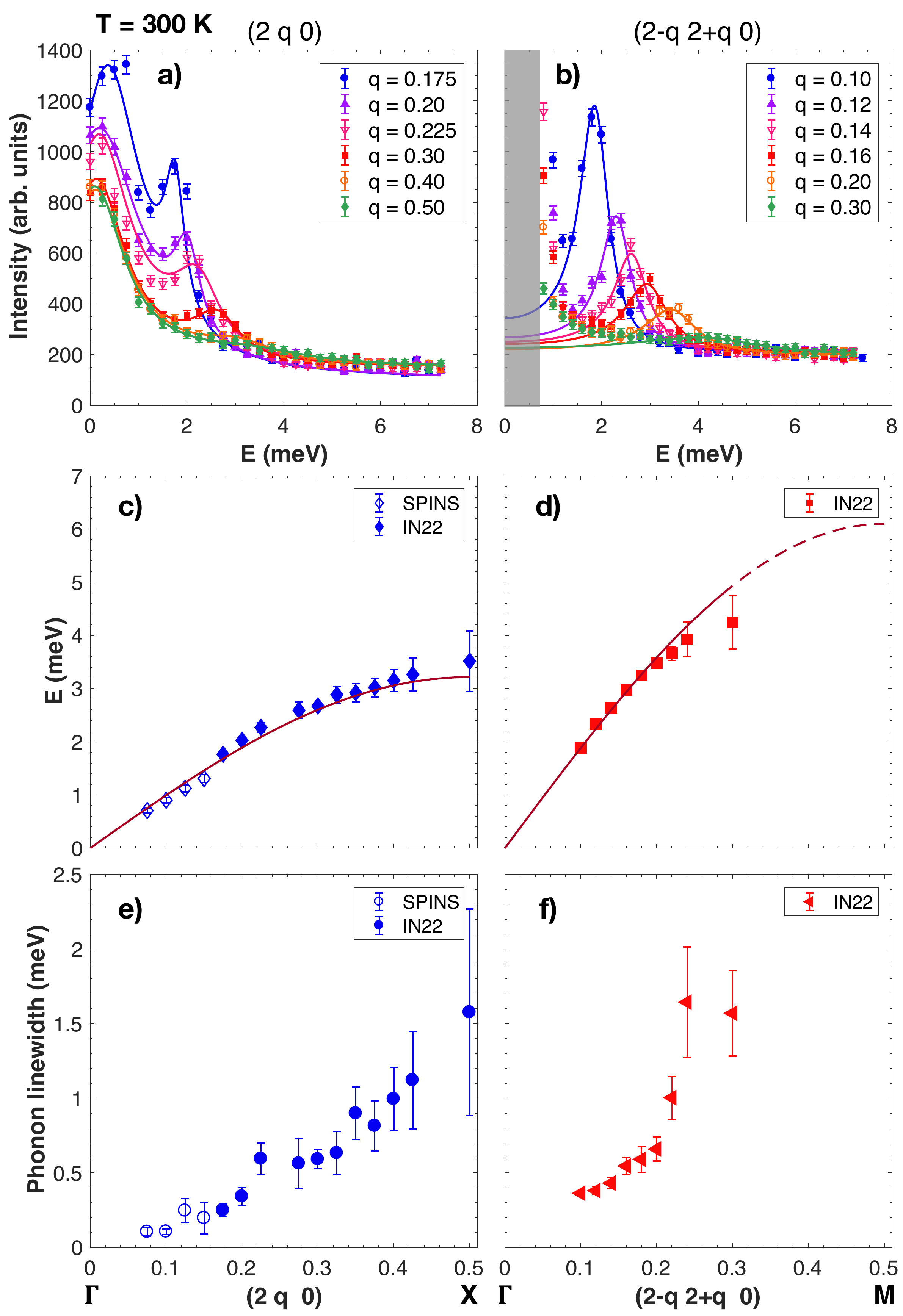}
 \caption{ \label{fig:dispersion} $(a)$-$(b)$ Constant-Q cuts through the acoustic phonon TA$_1$ for several Q positions from the (2~0~0) Bragg peak towards the $X$ point and from the (2~2~0) Bragg peak towards the $M$ point (TA$_2$ mode), respectively. The grey area corresponds to the region in energy contaminated by strong scattering from aluminium. $(c)$-$(d)$ Dispersion curves associated to the TA$_1$ mode towards the $X$ and the TA$_2$ mode towards the $M$ point respectively. The red line is a fit to a sine function. $(e)$-$(f)$  Q-dependence of the TA$_1$ and TA$_2$ phonon linewidths extracted from the constant-Q cuts as described in the text.}
\end{figure}

We first investigate the acoustic phonon dispersions with the aim of identifying where in reciprocal space any anomalies occur.  The phonon dispersions at room temperature were mapped out in the (H~K~0) plane.  The transverse acoustic phonon dispersions were measured around the (2~0~0) and (2~2~0) Bragg positions, in the [2~$q$~0] and [2-$q$~2+$q$~0] directions, from the $\Gamma$ zone center point towards the $X$ and $M$ Brillouin zone boundary symmetry points respectively, at 300~K on the thermal triple-axis IN22 (ILL). The measurements were extended to low momentum transfer, to approach the ${q\rightarrow 0}$ limit, on the cold triple axis SPINS (NIST), in the [2 $q$ 0] direction with the [2-$q$~2+$q$~0] not measurable with cold neutrons due to kinematic constraints of neutron scattering. Fig. \ref{fig:dispersion} $(a)$ and $(b)$ show constant-Q cuts around the (2~0~0) and (2~2~0) positions respectively, performed on IN22. A broad incoherent signal centred around E~=~0~meV can be observed and is attributed to incoherent scattering of the hydrogen atoms present in the MA molecules. On top of this background, clear acoustic phonon modes can be seen which are correlated and form well defined peaks in scans as a function of both momentum and energy transfer. 

The energy position $\omega_0$ of the harmonic phonon modes was extracted as a function of $q$ by fitting the experimental data (after subtracting a constant background) using a damped harmonic oscillator model:
 
 \begin{eqnarray}
 S(\vec{Q},\omega) = ... \nonumber \\
  \left[1 + n(\omega)\right]I_0 \left(\frac{\Gamma_0}{\Gamma_0^2 + (\omega-\omega_0)^2} -  \frac{\Gamma_0}{\Gamma_0^2 + (\omega + \omega_0)^2}\right) \nonumber
 \end{eqnarray}

\noindent where $\left[1 + n(\omega)\right]$ is the Bose factor, $I_0$ is a constant, and $\Gamma_0$ is the phonon energy linewidth inversely proportional to the phonon lifetime via $\Gamma_0 \sim {1\over \tau}$.  The above form of the neutron scattering cross section takes into account both neutron energy gain and loss and obeys detailed balance.~\cite{Shirane2004}  The incoherent part was fitted using the sum of a delta function centred around E = 0 meV convoluted with the instrumental resolution and a Lorentzian. In the case of the TA$_2$ modes, the incoherent part was ignored as the elastic signal was dominated by scattering coming from aluminium present in the sample environment.  

Fig. \ref{fig:dispersion} $(c)$-$(d)$ shows the resulting dispersion curves from $\Gamma$ to $X$ (TA$_1$ mode) and $\Gamma$ to $M$ (TA$_2$ mode) respectively.  Near the (2, 0, 0) and (2, 2, 0) zone centers, the phonons were found to be well defined both in energy and momentum.  While long wavelength acoustic phonons are associated with a uniform center of mass motion of the lattice~\cite{Dove1993,Shirane2004}, our results on MAPCl$_3$ imply long lived and spatially correlated lattice dynamics and therefore we associate these phonons with primarily the PbCl$_{3}$ framework instead of the interstitial molecules which are disordered, uncorrelated,  and fluctuating with a large energy scale at 300 K.  The molecular motions are expected to give rise to a quasielastic signal centered around the elastic (E=0 meV) line instead of dispersive harmonic modes well defined both in energy and momentum.~\cite{Bee2004}   We also note that these phonon modes are very close in energy to that observed in CsPbCl$_{3}$ inorganic variant~\cite{Fujii1974}.  Both TA$_{1}$ and TA$_{2}$ dispersion curves were fitted to a sine function allowing us to extract the elastic coefficient $C_{44}$ = 3.0(2) GPa from the TA$_1$ mode and ($C_{11}-C_{12}$)/2 = 10.8(1) GPa from the TA$_2$ mode. These values are consistent with the elastic constants reported in the iodine and bromine counterparts \cite{Letoublon2016, Ferreira2018,Beecher2016} as the chlorine compound is expected to be mechanically ``softer". The elastic coefficients extracted from neutron scattering for the three compounds are summarized in Table \ref{tab:coeff}. 

The phonon linewidth was also determined as a function of $q$ and shows a strong energy damping for momentum transfers approaching the zone boundaries, indicative of a shorter phonon lifetime for these smaller wavelength excitations. In particular, Fig. \ref{fig:dispersion} $(d)$ shows that the TA$_2$ mode significantly broadens in energy and cannot be observed definitively for q $>$ 0.3, close to the $M$ symmetry point, revealing a strong anharmonicity of this acoustic phonon mode for momentum transfers near the Brillouin zone boundary. Furthermore, the TA$_2$ dispersion shows that the phonon branch becomes flat towards the zone boundary, for q $\geq$ 0.3. 

While well defined phonon modes were observed throughout the Brillouin zone in the purely inorganic analogue CsPbCl$_3$ \cite{Fujii1974}, a measurable broadening in energy was observed near the zone boundaries and attributed to condensation of optic modes driving a structural transition.  The marked increase in linewidth  near the zone boundary in CsPbCl$_{3}$ is analogous to what we observe here in MAPbCl$_{3}$, although the dampening in the organic-inorganic variant is much stronger and more pronounced.   Although we attribute these acoustic phonon modes to the dynamics of the PbCl$_3$ framework, the organic cation may strongly affect these modes through hydrogen bonding between the molecule and inorganic framework as suggested based on powder diffraction studies of the structure~\cite{Chi2005} and also first principles calculations~\cite{Brivio2015}.  The broadening may originate from this coupling and the fast molecular dynamics reported based on neutron spectroscopy on powders of the Bromine variant.~\cite{Swainson2015}  This may explain why in MAPbCl$_3$ the TA$_1$ mode shows a strong damping towards the $X$ point while this was not the reported in the inorganic perovskite compound \cite{Fujii1974} and also why the broadening is more pronounced in this organic-inorganic perovskite.

\begin{table*}[t]
\caption{\label{tab:coeff}Comparison of elastic constants extracted from neutron inelastic measurements at 300 K of MAPbCl$_3$ with the MAPbI$_3$ and MAPbBr$_3$ counterparts.}
\begin{ruledtabular}
\begin{tabular}{llll}
& MAPbI$_3$ (from \cite{Ferreira2018}) & MAPbBr$_3$ (from \cite{Ferreira2018}) & MAPbCl$_3$ (this work)\\
 \hline
 $C_{44}$ (GPa)  & 7.3  & 4.1 & 3.0(2) \\
 ($C_{11}-C_{12}$)/2 (GPa) & 5.25 & 8 & 10.8(1) \\
\end{tabular}
\end{ruledtabular}
\end{table*}

\section{Temperature dependence}

\begin{figure}
 \includegraphics[scale=0.34]{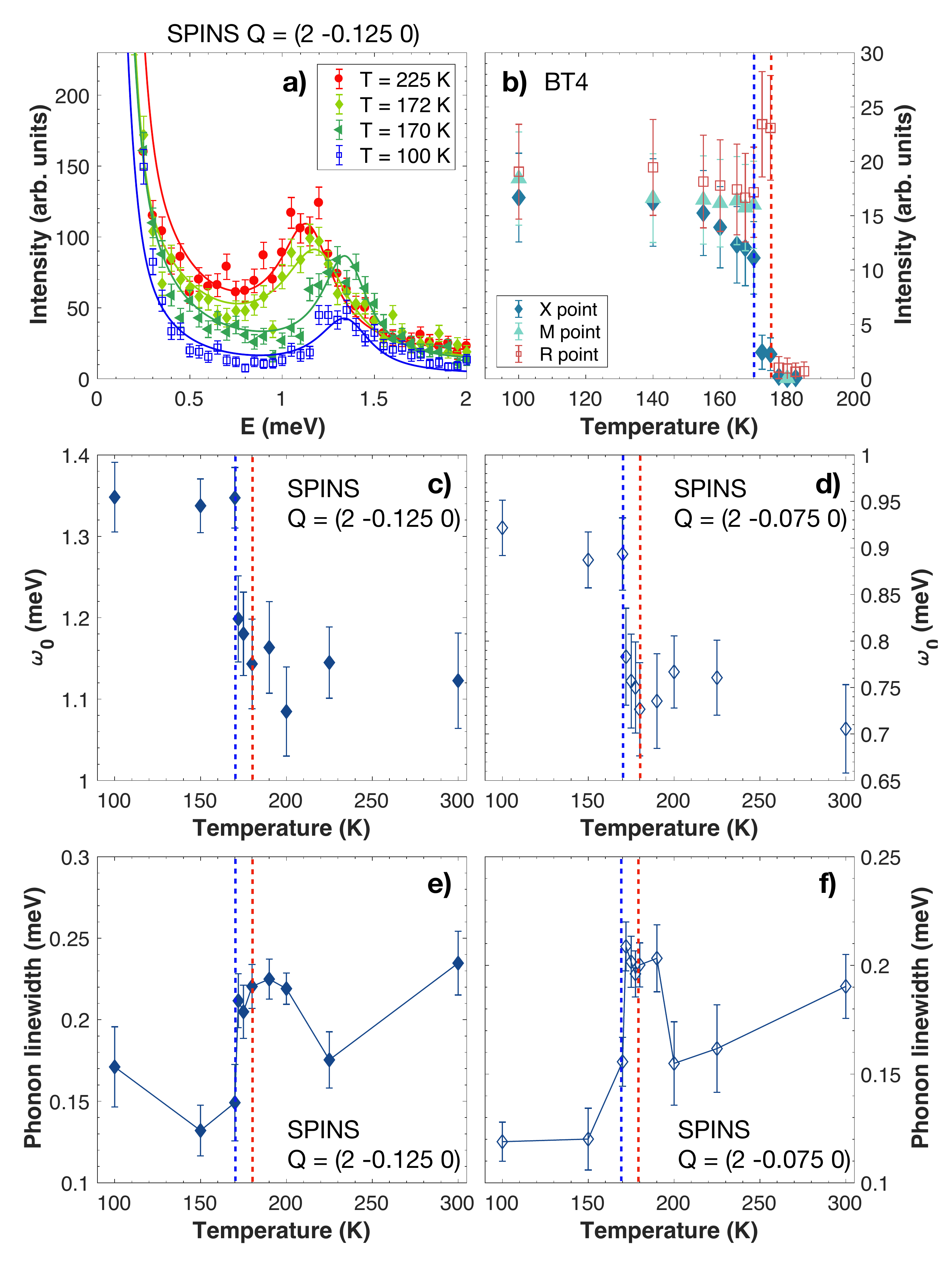}
 \caption{ \label{fig:evol_temp} $(a)$ Constant Q = (2~-0.125~0) scans at several temperatures, measured on SPINS (NIST). $(b)$ Temperature evolution of the integrated intensity measured with E = 0 at the $X$, $M$ and $R$ points measured on BT4 (NIST). $(c)$ - $(d)$ Temperature evolution of the energy position of the TA$_1$ mode, extracted at the Q = (2~-0.125~0) and Q = (2~-0.075~0) positions, respectively. $(e)$-$(f)$ Temperature evolution of the TA$_1$ phonon linewidth, extracted at the Q = (2~-0.125~0) and Q = (2~-0.075~0) positions, respectively.}
\end{figure}

Having identified strong phonon broadening near the $X$ and $M$ zone boundaries at 300 K, we now discuss the temperature dependence.  In order to characterize the structural phase transitions in MAPbCl$_3$, the cell parameter $a$ (Fig. \ref{fig:structure} $c)$ and the intensity of elastic (E=0 meV) superlattice Bragg reflections (Fig. \ref{fig:evol_temp} $b$) at the $X$, $M$ and $R$ zone boundaries (Fig. \ref{fig:structure} $a$) were measured as a function of temperature, using the thermal triple-axis spectrometer BT4 (NIST).  At low temperature, new nuclear and momentum resolution-limited Bragg peaks were found at all three zone boundaries implying a doubling of the unit cell along the crystallographic directions on entering the low temperature orthorhombic phase from the high temperature cubic phase.  As shown in Fig. \ref{fig:structure} $(c)$, the cell parameter evolution shows a sudden jump around 177 K followed by a cusp around 170 K, to the low temperature orthorhombic phase. Although the cell parameter behavior is different for each transition, the heat capacity measurements display two similar anomalies, and thus the nature of each transition (i.e. whether the transition is first or second order) cannot be conclusively determined from these results. 

Fig. \ref{fig:evol_temp} $(b)$ shows the temperature dependence of the elastic intensities at $(2,\frac{1}{2}, 0)$ ($X$ point), $(\frac{3}{2},\frac{1}{2}, 0)$ ($M$ point) and $(\frac{3}{2},\frac{3}{2},\frac{5}{2})$ ($R$ point). Both $X$ and $R$ point show abrupt changes at 170 K and 177 K. On entering the tetragonal phase from the high temperature cubic phase, the intensity of new nuclear Bragg peaks at the $R$ and $X$ point increases.  A further anomaly is observed on entering the orthorhombic phase with the onset of a nuclear Bragg peak at the $M$ point, while the intensity for the $X$ and $R$ points become comparable.  While the $X$  $(2,\frac{1}{2}, 0)$ and $R$ $(\frac{3}{2},\frac{3}{2},\frac{5}{2})$ positions are distinct in the cubic phase, these positions are the same in the tetragonal phase.  The equivalence explains why the Bragg peaks at these positions respond in concert on entering the tetragonal phase and then both show further simultaneous changes on entering the low temperature orthorhombic phase.  We discuss the implications of these zone boundaries on the lattice distortion below in the discussion section.

Having shown the temperature dependence of the static neutron response, we now present the temperature dependence of the dynamics.  To track the long wavelength dynamics near the Brillouin zone center $\Gamma$, several chosen constant-$Q$ cuts through the acoustic phonon mode TA$_1$ were measured as a function of temperature on SPINS, from room temperature down to 100~K. Figures \ref{fig:evol_temp} $(c)$-$(f)$ display the temperature evolution of the energy position and the phonon linewidth, extracted from constant Q~=~(2~-0.125~0) (Fig. \ref{fig:evol_temp} $a$ and $c$) and Q~=~(2~-0.075~0) (panel $d$) scans. A sudden change of the acoustic branch can be observed when the system approaches the structural transitions on increasing temperature from the orthorhombic phase, as a significant drop in the energy position occurs between 170 K and 180 K (panels $c-d$), although the energy of the phonon mode remains finite up to 300~K at all wavevectors measured away from the $\Gamma$ point. Constant-Q cuts were also measured at higher Q on IN22 as a function of temperature and also show an energy softening when going toward higher temperatures, indicating that the anomaly in energy position is not only located near the zone center but affects the entire phonon branch. However, we note that a similar discontinuity was not clearly observed for the TA$_2$ mode.   We emphasize that this is not a soft mode, and within the resolution of our measurements this is not confined to a particular range of wavevectors (like in Jahn Teller transitions such as discussed in Ref. \onlinecite{Weber2017}), but rather a discontinuity associated with the structural transition and hence the elastic constants.

We now present data addressing the temperature dependence of linewidth of the TA$_1$ mode. As shown in Fig. \ref{fig:evol_temp} $(e)$ and $(f)$, the phonon linewidth decreases on cooling but reaches a maximum at temperatures just above the phase transition. This anomaly in the linewidth, observed for several values of $q$, accompanies the hardening of the phonon mode on cooling. This feature was not reported in the CsPbCl$_3$ inorganic compound  \cite{Fujii1974} and is discussed below in the context of the dynamics previously reported in SrTiO$_{3}$.   

\begin{figure}
 \includegraphics[scale=0.42]{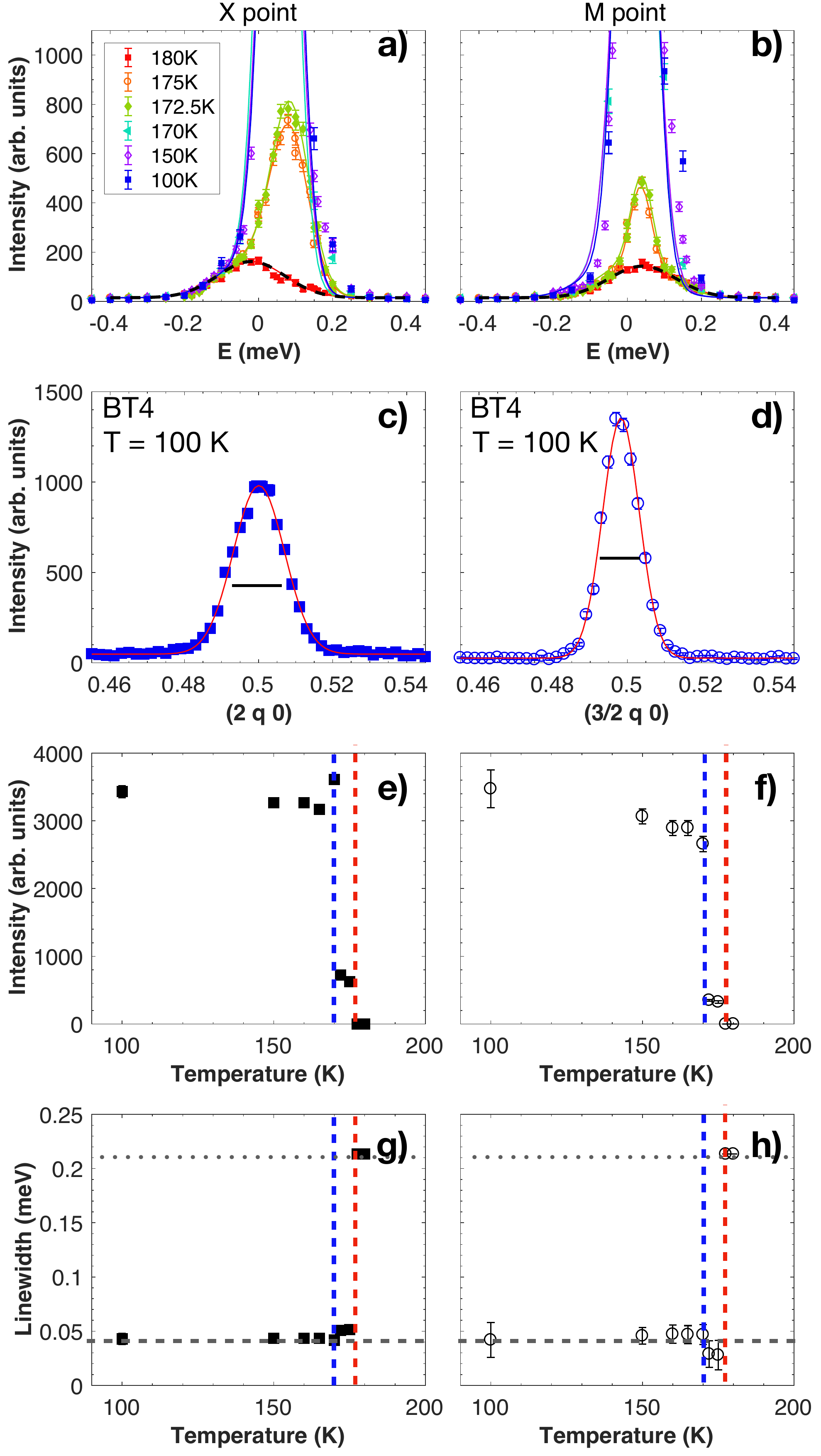}
 \caption{ \label{fig:critical_scat} $(a)-(b)$ Energy scans measured on TASP (PSI) around $X = (2,\frac{1}{2}, 0)$ and $M = (\frac{3}{2},\frac{1}{2}, 0)$ respectively, at several temperatures. $(c)-(d)$ Momentum scans around the $X$ and $M$ points performed on BT4 at T = 100 K. The solid lines show the instrumental Q resolution. $(e)-(f)$ Temperature evolution of the integrated intensity extracted from these energy scans. $(g)-(h)$ Temperature evolution of the linewidth extracted from the energy scans. The grey dotted lines show the energy resolution extracted from the width of the incoherent line of vanadium, the grey dashed lines show the Bragg width and the colored dashed lines limit the two structural transitions of MAPbCl$_3$.}
\end{figure}

In parallel to the long wavelength acoustic fluctuations associated to the inorganic framework measured near the zone center $\Gamma$ point, energy scans around the E=0 elastic line at the $X$ and $M$ zone boundaries were followed as a function of temperature, using the cold triple-axis TASP (PSI), and are summarized in Fig. \ref{fig:critical_scat}.   Fig. \ref{fig:critical_scat} $(a-b)$ shows energy scans at $(2,\frac{1}{2}, 0)$ ($X$ point), $(\frac{3}{2},\frac{1}{2}, 0)$ ($M$ point) for several temperatures.  The data illustrates  a dramatic increase in intensity with the width in energy being resolution limited for all temperatures.  Fig. \ref{fig:critical_scat} $(c-d)$ display momentum scans taken on the BT4 spectrometer at the elastic line showing that this increase in intensity at the zone boundary is tied to a resolution limited peak in momentum.  Therefore, the increase in scattering is not due to incoherent scattering from hydrogen which would not show such a structure in momentum transfer but would rather give rise to a momentum broadened and featureless response. Moreover, we do not observe any evidence of dynamic fluctuations nor any broadening of correlations in momentum that would indicate a finite correlation length. Associated with the zone boundary phase transitions, we could not observe any slowing of fluctuations in energy or momentum that would indicate dynamic critical behavior beyond the resolution of the spectrometer~\cite{Collins1989}, but only the increase in intensity of a peak which is resolution limited in momentum and energy as the temperature is decreased.  

To parameterize the observed neutron scattering cross section, the experimental data was fitted to the sum of a $\delta$-function centered around E = 0 meV, convoluted with the instrumental resolution (shown with dashed black lines) and a Gaussian function. The offset in the energy position of the peaks is explained by a change of lattice parameter with temperature, and therefore an offset in $\vec{Q}$ of the maximum of intensity. As presented on Fig. \ref{fig:critical_scat}, while the intensity shows a sudden increase at the 170~K transition (panels $e$ and $f$) for both $X$ and $M$ positions, the width in energy of the low temperature zone boundary peaks (half width at half maximum) drops and becomes less than the resolution extracted from the width of the incoherent elastic line. This sharp feature has an energy width that is associated with a delta function in momentum and energy. Its width therefore corresponds to the Bragg resolution of the instrument, which is consequently sharper than the energy resolution extracted from a momentum broadened incoherent signal like that measured from an incoherent scattering vanadium standard.~\cite{Shirane1967} The appearance of these momentum and energy resolution limited peaks can be understood as the onset of superlattice Bragg peaks when going towards the orthorhombic transition, as the unit cell is doubled in all three crystallographic directions. 

\section{Discussion}

\subsubsection{Temperature dependence of order parameters and phonon softening}

In the purely inorganic counterparts CsPbCl$_3$ and CsPbBr$_3$ \cite{Fujii1974, Hua1991, Hirotsu1974}, the higher temperature transition is accompanied by the appearance of superlattice Bragg reflections at the $M$ point followed by superlattice reflection appearing at the $X$ and $R$ points towards the lower temperature transition.  In the tetragonal phase, just as is the case for the organic-inorganic variant discussed above, the $X$ and $R$ reciprocal lattice points are equivalent.  Therefore, the higher temperature transition is expected to be driven by the softening of phonon modes at the $M$ zone boundary while the lower temperature transition should be caused by the softening of phonon modes at the $R$ zone boundary, as supported by symmetry analysis.  New Bragg peaks at the zone boundaries was found to be accompanied by dynamic fluctuations in energy indicative of a diverging slowing down of inelastic critical dynamics.

In contrast to these purely inorganic materials, our neutron study shows that organic-inorganic MAPbCl$_3$ displays a different temperature evolution of superlattice reflections as the intensity at the $X$ and $R$ points appear first when cooling from the cubic phase to the tetragonal phase followed by the $M$ position at low temperatures on entering the orthorhombic phase. As noted in Ref. \onlinecite{Fujii1974} in the context of CsPbCl$_{3}$ and further discussed in Ref. \onlinecite{Lynn1978} in relation to neutron scattering structure factors, the $M$ point characterizes a rigid tilt distortion, while the $R$ point represents a displacive distortion of the octahedra. While in the inorganic CsPbCl$_{3}$, a rigid tilt distortion occurs at high temperatures followed by a displacive distortion of the octahedra on cooling, the inorganic-organic MAPbCl$_3$ undergoes a different sequence of transitions, with the displacive distortion of the octahedra occurring at high temperatures followed by the tilt distortion at lower temperatures.  The results highlight the importance of condensation of displacive phonons over tilt distortions~\cite{Ghosh2017} in MAPbCl$_3$, possibly caused by hydrogen bonding~\cite{Lee2015,Lee2016}, as the molecular motions freeze near the ordering transitions.~\cite{Chi2005}  The combination of both a tilt and displacive transition is consistent with low temperature neutron diffraction results.~\cite{Swainson2003}

We now discuss the temperature behavior of the TA$_1$ mode near the $\Gamma$ zone centre point. While in most oxide perovskites one would expect the structural transitions to be driven by the softening of associated phonon modes (as was reported in SrTiO$_3$ \cite{Shirane1969, Shapiro1972}), MAPbCl$_3$ does not show a progressive softening of the acoustic phonon when approaching the low-temperature transition from below but rather a sudden drop in the acoustic phonon frequencies along the entire branch on heating through the structural transition. Moreover, when crossing the higher temperature transition, these frequencies remain constant within experimental error up to $\sim$ 300~K. This illustrates that simultaneously with the zone boundary distortions discussed above, the elastic $C_{44}$ constant characterized by the slope of the acoustic branch in the $lim_{q\rightarrow 0}$ also responds to the structural transition characterized by new Bragg peaks at the zone boundaries.  This is suggestive of a coupling between zone center and boundary dynamics.

In the diffraction study of the ordered phase of MAPbCl$_3$ \cite{Chi2005}, L. Chi \textit{et al.} point out that the structure is strongly distorted, due to the strong hydrogen bonding between methylammonium and chloride ions. Indeed, as the chlorine counterpart has the smallest radius in the halide family, the decreased size of the cage strengthens the hydrogen bonding, causing the molecular cation to be the more rigid unit in the system. Both the distortion of octahedra and the orientational ordering of the molecule become thus incompatible and the system is consequently forced to distort when crossing the low temperature transition. Hence this may explain the experimental absence with neutron scattering of a proper E=0 softening, of the phonon modes in MAPbCl$_3$ and that the transition is better described in terms of an order-disorder nature as found in molecular systems such as calcite~\cite{Hagen1992} and discussed below. It is worth noting that this contrasts with the bromine counterpart, which shows a clear phonon softening \cite{Swainson2015}, as the tilting of octahedra is more significant than in the chlorine compound.  Finally, it was pointed out by Y. Fujii \textit{et al.} and S. Hirotsu \textit{et al.}  \cite{Fujii1974, Hirotsu1974} that the expected soft modes in CsPbCl$_3$ and CsPbBr$_3$ could not be observed and that only dynamic critical inelastic scattering, outside the spectrometer resolution, around the $R$ and $M$ points could be measured through the phase transitions.

\subsubsection{Coupling between molecular cations and inorganic framework}

The presence of strong hydrogen bonding between the methylammonium cation and the chlorine ions raises the issue of a coupling between the inorganic framework and the molecule. We discuss in the following the experimental evidence in our study for such coupling.  In particular, our experiment finds that the structural transition in MAPbCl$_3$ is not characterized by a critical slowing down of fluctuations, but rather a momentum and energy resolution limited peak.

Structural transitions can usually be classified~\cite{Cowley1980}  in either displacive\cite{Cowley1976, Bechtel2018,Shirane1974, Scott1974} or order-disorder in nature. Examples of displacive transitions correspond to PbTiO$_{3}$~\cite{Shirane1970,Kempa2006} or BaTiO$_{3}$~\cite{Shirane1967} and representative order-disorder transitions include ammonium halides such as NH$_4$Cl or NH$_4$Br \cite{Yamada1974, Yamada1972}.  This later class of transitions is often governed by two timescales, one relatively short time scale being set by the soft-phonon mode and one longer, critical timescale associated with the new elastic Bragg peak and models have been developed to describe the scattering cross section for this. ~\cite{Halperin1976,Aubry1975,Bruce1980bis, Bruce1980,Huber1974}  

All of these models aim at providing a unified approach to best describe the lattice dynamics through structural transitions and are built around the general idea of a coupling between components with two different time scales. The particular structure of hybrid organic-inorganic perovskites is specifically suited to understand this coupling. At high temperature, both the inorganic framework and the molecular host develop fast motions around an equilibrium position. They also form two sublattices which are coupled through hydrogen bondings. As the molecular host is disordered in the high temperature phase, the excitation spectrum is dominated by harmonic phonons, primarily associated with the distortion of the PbCl$_6$ octahedra, although the fast motion of the molecules may cause the damping of the phonons near the zone boundaries. Approaching the structural transition to the orthorhombic phase, the thermal fluctuations decrease and the symmetry breaking corresponding to the new structure allows the molecular host to order in a preferred position with local ordering.~\cite{Comin2016}  In particular, Chi \textit{et al.} \cite{Chi2005} showed that in the ordered phase, the CH$_3$NH$_3$ cations align in an antiparallel arrangement in the chlorine compound. Thus, right above the transition, the molecular host develops slower dynamics corresponding to jumps from a preferred orientation to another in addition to fast motions around these equilibrium positions.  These two timescales near the structural transition has been observed with high resolution backscattering neutron spectroscopy in the Br variant.~\cite{Brown2017}  The slower dynamics can be associated to the appearance of the static Bragg peaks at the zone boundary, indicating the growing coherence of molecular motions and precursor of superlattice Bragg peaks. However in this case it should be noted that the critical narrowing could not be captured by using a cold triple-axis spectrometer, and further high-resolution measurements may be necessary to probe a wider dynamic range and check for the existence of dynamic critical scattering around the new Bragg peaks in the inelastic channel\cite{Stock2010}.

The presence of a second timescale is corroborated by the enhanced broadening of the acoustic phonons near the structural transitions as illustrated in Fig. \ref{fig:evol_temp} in panels $e$ and $f$.  This increase in broadening is indicative of a second energy scale and coupling between the slowing molecular dynamics and the acoustic fluctuations of the framework lattice, through the strong hydrogen bondings, which provide a bridge between the two dynamics.  Such an increase has been predicted by theories involving structural transitions with two timescales applied to SrTiO$_{3}$.~\cite{Halperin1976}  A key difference in the case of MAPCl$_{3}$ is that we do not observe an optical soft mode with our resolution, however inelastic x-ray results have suggested the presence of a mode at higher energies.~\cite{Comin2016}

In conclusion, this work reports the presence of acoustic phonon energy broadening in MAPbCl$_3$ when going through structural transitions. A detailed temperature dependence study is presented and shows the appearance of a new Bragg peaks at the zone boundary when decreasing the temperature, along with a hardening of the TA$_1$ phonon mode and an anomaly in the phonon linewidth at the tetragonal-to-orthorhombic transition. This is interpreted as evidence of the coupling between the acoustic TA$_1$ phonon mode and molecular reorientation, thanks to strong hydrogen bonding.

\begin{acknowledgments}
We acknowledge funding from the EPSRC and the STFC. We are thankful to P. Bourges and I. P. Swainson for fruitful discussions, and T. Fennell for measurements on EIGER (PSI). We acknowledge the support of the National Institute of Standards and Technology, U.S. Department of Commerce, in providing the neutron research facilities used in this work.

\end{acknowledgments}

\bibliography{MAPCl_rev2}

\begin{thebibliography}{53}%
\makeatletter
\providecommand \@ifxundefined [1]{%
 \@ifx{#1\undefined}
}%
\providecommand \@ifnum [1]{%
 \ifnum #1\expandafter \@firstoftwo
 \else \expandafter \@secondoftwo
 \fi
}%
\providecommand \@ifx [1]{%
 \ifx #1\expandafter \@firstoftwo
 \else \expandafter \@secondoftwo
 \fi
}%
\providecommand \natexlab [1]{#1}%
\providecommand \enquote  [1]{``#1''}%
\providecommand \bibnamefont  [1]{#1}%
\providecommand \bibfnamefont [1]{#1}%
\providecommand \citenamefont [1]{#1}%
\providecommand \href@noop [0]{\@secondoftwo}%
\providecommand \href [0]{\begingroup \@sanitize@url \@href}%
\providecommand \@href[1]{\@@startlink{#1}\@@href}%
\providecommand \@@href[1]{\endgroup#1\@@endlink}%
\providecommand \@sanitize@url [0]{\catcode `\\12\catcode `\$12\catcode
  `\&12\catcode `\#12\catcode `\^12\catcode `\_12\catcode `\%12\relax}%
\providecommand \@@startlink[1]{}%
\providecommand \@@endlink[0]{}%
\providecommand \url  [0]{\begingroup\@sanitize@url \@url }%
\providecommand \@url [1]{\endgroup\@href {#1}{\urlprefix }}%
\providecommand \urlprefix  [0]{URL }%
\providecommand \Eprint [0]{\href }%
\providecommand \doibase [0]{http://dx.doi.org/}%
\providecommand \selectlanguage [0]{\@gobble}%
\providecommand \bibinfo  [0]{\@secondoftwo}%
\providecommand \bibfield  [0]{\@secondoftwo}%
\providecommand \translation [1]{[#1]}%
\providecommand \BibitemOpen [0]{}%
\providecommand \bibitemStop [0]{}%
\providecommand \bibitemNoStop [0]{.\EOS\space}%
\providecommand \EOS [0]{\spacefactor3000\relax}%
\providecommand \BibitemShut  [1]{\csname bibitem#1\endcsname}%
\let\auto@bib@innerbib\@empty
\bibitem [{\citenamefont {Fujii}\ \emph {et~al.}(1974)\citenamefont {Fujii},
  \citenamefont {Hoshino}, \citenamefont {Yamada},\ and\ \citenamefont
  {Shirane}}]{Fujii1974}%
  \BibitemOpen
  \bibfield  {author} {\bibinfo {author} {\bibfnamefont {Y.}~\bibnamefont
  {Fujii}}, \bibinfo {author} {\bibfnamefont {S.}~\bibnamefont {Hoshino}},
  \bibinfo {author} {\bibfnamefont {Y.}~\bibnamefont {Yamada}}, \ and\ \bibinfo
  {author} {\bibfnamefont {G.}~\bibnamefont {Shirane}},\ }\href@noop {}
  {\bibfield  {journal} {\bibinfo  {journal} {Phys. Rev. B}\ }\textbf {\bibinfo
  {volume} {9}},\ \bibinfo {pages} {4549} (\bibinfo {year} {1974})}\BibitemShut
  {NoStop}%
\bibitem [{\citenamefont {Saparov}\ and\ \citenamefont
  {Mitzi}(2016)}]{Saparov2016}%
  \BibitemOpen
  \bibfield  {author} {\bibinfo {author} {\bibfnamefont {B.}~\bibnamefont
  {Saparov}}\ and\ \bibinfo {author} {\bibfnamefont {D.~B.}\ \bibnamefont
  {Mitzi}},\ }\href@noop {} {\bibfield  {journal} {\bibinfo  {journal} {Chem.
  Rev.}\ }\textbf {\bibinfo {volume} {116}},\ \bibinfo {pages} {4558} (\bibinfo
  {year} {2016})}\BibitemShut {NoStop}%
\bibitem [{\citenamefont {Brenner}\ \emph {et~al.}(2016)\citenamefont
  {Brenner}, \citenamefont {Egger}, \citenamefont {Kronik}, \citenamefont
  {Hodes},\ and\ \citenamefont {Cahen}}]{Brenner2016}%
  \BibitemOpen
  \bibfield  {author} {\bibinfo {author} {\bibfnamefont {T.~M.}\ \bibnamefont
  {Brenner}}, \bibinfo {author} {\bibfnamefont {D.~A.}\ \bibnamefont {Egger}},
  \bibinfo {author} {\bibfnamefont {L.}~\bibnamefont {Kronik}}, \bibinfo
  {author} {\bibfnamefont {G.}~\bibnamefont {Hodes}}, \ and\ \bibinfo {author}
  {\bibfnamefont {D.}~\bibnamefont {Cahen}},\ }\href@noop {} {\bibfield
  {journal} {\bibinfo  {journal} {Nature Reviews Materials}\ }\textbf {\bibinfo
  {volume} {1}} (\bibinfo {year} {2016})}\BibitemShut {NoStop}%
\bibitem [{\citenamefont {Charles}\ \emph {et~al.}(2017)\citenamefont
  {Charles}, \citenamefont {Dillon}, \citenamefont {Weber}, \citenamefont
  {Islam},\ and\ \citenamefont {Weller}}]{Charles2017}%
  \BibitemOpen
  \bibfield  {author} {\bibinfo {author} {\bibfnamefont {B.}~\bibnamefont
  {Charles}}, \bibinfo {author} {\bibfnamefont {J.}~\bibnamefont {Dillon}},
  \bibinfo {author} {\bibfnamefont {O.~J.}\ \bibnamefont {Weber}}, \bibinfo
  {author} {\bibfnamefont {M.~S.}\ \bibnamefont {Islam}}, \ and\ \bibinfo
  {author} {\bibfnamefont {M.~T.}\ \bibnamefont {Weller}},\ }\href@noop {}
  {\bibfield  {journal} {\bibinfo  {journal} {J. Mater. Chem. A}\ }\textbf
  {\bibinfo {volume} {5}} (\bibinfo {year} {2017})}\BibitemShut {NoStop}%
\bibitem [{\citenamefont {Ghosh}\ \emph {et~al.}(2017)\citenamefont {Ghosh},
  \citenamefont {Atkins}, \citenamefont {Islam}, \citenamefont {B.},\ and\
  \citenamefont {Earnes}}]{Ghosh2017}%
  \BibitemOpen
  \bibfield  {author} {\bibinfo {author} {\bibfnamefont {D.}~\bibnamefont
  {Ghosh}}, \bibinfo {author} {\bibfnamefont {P.~W.}\ \bibnamefont {Atkins}},
  \bibinfo {author} {\bibfnamefont {M.~S.}\ \bibnamefont {Islam}}, \bibinfo
  {author} {\bibfnamefont {W.~A.}\ \bibnamefont {B.}}, \ and\ \bibinfo {author}
  {\bibfnamefont {C.}~\bibnamefont {Earnes}},\ }\href@noop {} {\bibfield
  {journal} {\bibinfo  {journal} {ACS Energy Lett.}\ }\textbf {\bibinfo
  {volume} {2}},\ \bibinfo {pages} {2424} (\bibinfo {year} {2017})}\BibitemShut
  {NoStop}%
\bibitem [{\citenamefont {Lee}\ \emph {et~al.}(2015)\citenamefont {Lee},
  \citenamefont {Bristowe}, \citenamefont {Bristowe},\ and\ \citenamefont
  {Cheetham}}]{Lee2015}%
  \BibitemOpen
  \bibfield  {author} {\bibinfo {author} {\bibfnamefont {J.}~\bibnamefont
  {Lee}}, \bibinfo {author} {\bibfnamefont {N.}~\bibnamefont {Bristowe}},
  \bibinfo {author} {\bibfnamefont {P.}~\bibnamefont {Bristowe}}, \ and\
  \bibinfo {author} {\bibfnamefont {A.}~\bibnamefont {Cheetham}},\ }\href
  {\doibase 10.1039/C5CC00979K} {\bibfield  {journal} {\bibinfo  {journal}
  {Chem. Comm.}\ }\textbf {\bibinfo {volume} {51}},\ \bibinfo {pages} {6434}
  (\bibinfo {year} {2015})}\BibitemShut {NoStop}%
\bibitem [{\citenamefont {Ren}\ \emph {et~al.}(2016)\citenamefont {Ren},
  \citenamefont {Oswald}, \citenamefont {Wang}, \citenamefont {T.},\ and\
  \citenamefont {Chan}}]{Ren2016}%
  \BibitemOpen
  \bibfield  {author} {\bibinfo {author} {\bibfnamefont {Y.}~\bibnamefont
  {Ren}}, \bibinfo {author} {\bibfnamefont {I.~W.}\ \bibnamefont {Oswald}},
  \bibinfo {author} {\bibfnamefont {X.}~\bibnamefont {Wang}}, \bibinfo {author}
  {\bibfnamefont {M.~G.}\ \bibnamefont {T.}}, \ and\ \bibinfo {author}
  {\bibfnamefont {J.~Y.}\ \bibnamefont {Chan}},\ }\href@noop {} {\bibfield
  {journal} {\bibinfo  {journal} {Cryst. Growth Des.}\ }\textbf {\bibinfo
  {volume} {16}},\ \bibinfo {pages} {2945} (\bibinfo {year}
  {2016})}\BibitemShut {NoStop}%
\bibitem [{\citenamefont {Weller}\ \emph {et~al.}(2015)\citenamefont {Weller},
  \citenamefont {Weber}, \citenamefont {Henry}, \citenamefont {Pumpo},\ and\
  \citenamefont {Hansen}}]{Weller2015}%
  \BibitemOpen
  \bibfield  {author} {\bibinfo {author} {\bibfnamefont {M.~T.}\ \bibnamefont
  {Weller}}, \bibinfo {author} {\bibfnamefont {O.~J.}\ \bibnamefont {Weber}},
  \bibinfo {author} {\bibfnamefont {P.~F.}\ \bibnamefont {Henry}}, \bibinfo
  {author} {\bibfnamefont {A.~M.}\ \bibnamefont {Pumpo}}, \ and\ \bibinfo
  {author} {\bibfnamefont {T.~C.}\ \bibnamefont {Hansen}},\ }\href@noop {}
  {\bibfield  {journal} {\bibinfo  {journal} {Chem. Comm.}\ }\textbf {\bibinfo
  {volume} {51}} (\bibinfo {year} {2015})}\BibitemShut {NoStop}%
\bibitem [{\citenamefont {Hirotsu}\ \emph {et~al.}(1974)\citenamefont
  {Hirotsu}, \citenamefont {Harada},\ and\ \citenamefont
  {Iizumi}}]{Hirotsu1974}%
  \BibitemOpen
  \bibfield  {author} {\bibinfo {author} {\bibfnamefont {S.}~\bibnamefont
  {Hirotsu}}, \bibinfo {author} {\bibfnamefont {J.}~\bibnamefont {Harada}}, \
  and\ \bibinfo {author} {\bibfnamefont {M.}~\bibnamefont {Iizumi}},\
  }\href@noop {} {\bibfield  {journal} {\bibinfo  {journal} {J. Phys. Soc.
  Jpn}\ }\textbf {\bibinfo {volume} {37}},\ \bibinfo {pages} {1393} (\bibinfo
  {year} {1974})}\BibitemShut {NoStop}%
\bibitem [{\citenamefont {Hua}(1991)}]{Hua1991}%
  \BibitemOpen
  \bibfield  {author} {\bibinfo {author} {\bibfnamefont {G.~L.}\ \bibnamefont
  {Hua}},\ }\href@noop {} {\bibfield  {journal} {\bibinfo  {journal} {J. Phys.:
  Condens. Matter}\ }\textbf {\bibinfo {volume} {3}},\ \bibinfo {pages} {1371}
  (\bibinfo {year} {1991})}\BibitemShut {NoStop}%
\bibitem [{\citenamefont {Egger}\ \emph {et~al.}(2016)\citenamefont {Egger},
  \citenamefont {Rappe},\ and\ \citenamefont {Kronik}}]{Egger2016}%
  \BibitemOpen
  \bibfield  {author} {\bibinfo {author} {\bibfnamefont {D.~A.}\ \bibnamefont
  {Egger}}, \bibinfo {author} {\bibfnamefont {A.~M.}\ \bibnamefont {Rappe}}, \
  and\ \bibinfo {author} {\bibfnamefont {L.}~\bibnamefont {Kronik}},\
  }\href@noop {} {\bibfield  {journal} {\bibinfo  {journal} {Acc. Chem. Res.}\
  }\textbf {\bibinfo {volume} {49}},\ \bibinfo {pages} {573} (\bibinfo {year}
  {2016})}\BibitemShut {NoStop}%
\bibitem [{\citenamefont {Brown}\ \emph {et~al.}(2017)\citenamefont {Brown},
  \citenamefont {Parker}, \citenamefont {Garcia}, \citenamefont {Mukhopadhyay},
  \citenamefont {Sakai},\ and\ \citenamefont {Stock}}]{Brown2017}%
  \BibitemOpen
  \bibfield  {author} {\bibinfo {author} {\bibfnamefont {K.~L.}\ \bibnamefont
  {Brown}}, \bibinfo {author} {\bibfnamefont {S.~F.}\ \bibnamefont {Parker}},
  \bibinfo {author} {\bibfnamefont {I.~R.}\ \bibnamefont {Garcia}}, \bibinfo
  {author} {\bibfnamefont {S.}~\bibnamefont {Mukhopadhyay}}, \bibinfo {author}
  {\bibfnamefont {V.~G.}\ \bibnamefont {Sakai}}, \ and\ \bibinfo {author}
  {\bibfnamefont {C.}~\bibnamefont {Stock}},\ }\href@noop {} {\bibfield
  {journal} {\bibinfo  {journal} {Phys. Rev. B}\ }\textbf {\bibinfo {volume}
  {96}},\ \bibinfo {pages} {174111} (\bibinfo {year} {2017})}\BibitemShut
  {NoStop}%
\bibitem [{\citenamefont {Swainson}\ \emph {et~al.}(2015)\citenamefont
  {Swainson}, \citenamefont {Stock}, \citenamefont {Parker}, \citenamefont
  {Van~Eijck}, \citenamefont {Russina},\ and\ \citenamefont
  {Taylor}}]{Swainson2015}%
  \BibitemOpen
  \bibfield  {author} {\bibinfo {author} {\bibfnamefont {I.~P.}\ \bibnamefont
  {Swainson}}, \bibinfo {author} {\bibfnamefont {C.}~\bibnamefont {Stock}},
  \bibinfo {author} {\bibfnamefont {S.~F.}\ \bibnamefont {Parker}}, \bibinfo
  {author} {\bibfnamefont {L.}~\bibnamefont {Van~Eijck}}, \bibinfo {author}
  {\bibfnamefont {M.}~\bibnamefont {Russina}}, \ and\ \bibinfo {author}
  {\bibfnamefont {J.~W.}\ \bibnamefont {Taylor}},\ }\href@noop {} {\bibfield
  {journal} {\bibinfo  {journal} {Phys. Rev. B}\ }\textbf {\bibinfo {volume}
  {92}},\ \bibinfo {pages} {100303(R)} (\bibinfo {year} {2015})}\BibitemShut
  {NoStop}%
\bibitem [{\citenamefont {Potglisch}\ and\ \citenamefont
  {Weber}(1987)}]{Potglisch1987}%
  \BibitemOpen
  \bibfield  {author} {\bibinfo {author} {\bibfnamefont {A.}~\bibnamefont
  {Potglisch}}\ and\ \bibinfo {author} {\bibfnamefont {D.}~\bibnamefont
  {Weber}},\ }\href@noop {} {\bibfield  {journal} {\bibinfo  {journal} {J.
  Chem. Phys.}\ }\textbf {\bibinfo {volume} {87}},\ \bibinfo {pages} {6373}
  (\bibinfo {year} {1987})}\BibitemShut {NoStop}%
\bibitem [{\citenamefont {Chi}\ \emph {et~al.}(2005)\citenamefont {Chi},
  \citenamefont {Swainson}, \citenamefont {Cranswick}, \citenamefont {Her},
  \citenamefont {Stephens},\ and\ \citenamefont {Knop}}]{Chi2005}%
  \BibitemOpen
  \bibfield  {author} {\bibinfo {author} {\bibfnamefont {L.}~\bibnamefont
  {Chi}}, \bibinfo {author} {\bibfnamefont {I.}~\bibnamefont {Swainson}},
  \bibinfo {author} {\bibfnamefont {L.}~\bibnamefont {Cranswick}}, \bibinfo
  {author} {\bibfnamefont {J.}~\bibnamefont {Her}}, \bibinfo {author}
  {\bibfnamefont {P.}~\bibnamefont {Stephens}}, \ and\ \bibinfo {author}
  {\bibfnamefont {O.}~\bibnamefont {Knop}},\ }\href@noop {} {\bibfield
  {journal} {\bibinfo  {journal} {J. Solid State Chem.}\ }\textbf {\bibinfo
  {volume} {178}},\ \bibinfo {pages} {1376} (\bibinfo {year}
  {2005})}\BibitemShut {NoStop}%
\bibitem [{\citenamefont {Kawamura}\ and\ \citenamefont
  {Mashiyama}(1999)}]{Kawamura1999}%
  \BibitemOpen
  \bibfield  {author} {\bibinfo {author} {\bibfnamefont {Y.}~\bibnamefont
  {Kawamura}}\ and\ \bibinfo {author} {\bibfnamefont {H.}~\bibnamefont
  {Mashiyama}},\ }\href@noop {} {\bibfield  {journal} {\bibinfo  {journal} {J.
  Korean Phy. Soc.}\ }\textbf {\bibinfo {volume} {35}},\ \bibinfo {pages}
  {S1437} (\bibinfo {year} {1999})}\BibitemShut {NoStop}%
\bibitem [{\citenamefont {Swainson}(2005)}]{Swainson2005}%
  \BibitemOpen
  \bibfield  {author} {\bibinfo {author} {\bibfnamefont {I.~P.}\ \bibnamefont
  {Swainson}},\ }\href@noop {} {\bibfield  {journal} {\bibinfo  {journal} {Acta
  Cryst.}\ }\textbf {\bibinfo {volume} {B61}},\ \bibinfo {pages} {616}
  (\bibinfo {year} {2005})}\BibitemShut {NoStop}%
\bibitem [{\citenamefont {Guo}\ \emph {et~al.}(2017)\citenamefont {Guo},
  \citenamefont {Yaffe}, \citenamefont {Paley}, \citenamefont {Beecher},
  \citenamefont {Hull}, \citenamefont {Szpak}, \citenamefont {Owen},
  \citenamefont {Brus},\ and\ \citenamefont {Pimenta}}]{Guo2017}%
  \BibitemOpen
  \bibfield  {author} {\bibinfo {author} {\bibfnamefont {Y.}~\bibnamefont
  {Guo}}, \bibinfo {author} {\bibfnamefont {O.}~\bibnamefont {Yaffe}}, \bibinfo
  {author} {\bibfnamefont {D.}~\bibnamefont {Paley}}, \bibinfo {author}
  {\bibfnamefont {A.}~\bibnamefont {Beecher}}, \bibinfo {author} {\bibfnamefont
  {T.}~\bibnamefont {Hull}}, \bibinfo {author} {\bibfnamefont {G.}~\bibnamefont
  {Szpak}}, \bibinfo {author} {\bibfnamefont {J.}~\bibnamefont {Owen}},
  \bibinfo {author} {\bibfnamefont {L.}~\bibnamefont {Brus}}, \ and\ \bibinfo
  {author} {\bibfnamefont {M.}~\bibnamefont {Pimenta}},\ }\href@noop {}
  {\bibfield  {journal} {\bibinfo  {journal} {Phys. Rev. Mater.}\ }\textbf
  {\bibinfo {volume} {1}},\ \bibinfo {pages} {042401(R)} (\bibinfo {year}
  {2017})}\BibitemShut {NoStop}%
\bibitem [{\citenamefont {Leguy}\ \emph {et~al.}(2015)\citenamefont {Leguy},
  \citenamefont {Frost}, \citenamefont {P.}, \citenamefont {Sakai},
  \citenamefont {Kockelmann}, \citenamefont {Law}, \citenamefont {Li},
  \citenamefont {Foglia}, \citenamefont {Walsh}, \citenamefont {C},
  \citenamefont {Nelson}, \citenamefont {Cabral},\ and\ \citenamefont
  {Barnes}}]{Leguy2015}%
  \BibitemOpen
  \bibfield  {author} {\bibinfo {author} {\bibfnamefont {A.~M.}\ \bibnamefont
  {Leguy}}, \bibinfo {author} {\bibfnamefont {J.~M.}\ \bibnamefont {Frost}},
  \bibinfo {author} {\bibfnamefont {M.~A.}\ \bibnamefont {P.}}, \bibinfo
  {author} {\bibfnamefont {V.~G.}\ \bibnamefont {Sakai}}, \bibinfo {author}
  {\bibfnamefont {W.}~\bibnamefont {Kockelmann}}, \bibinfo {author}
  {\bibfnamefont {C.}~\bibnamefont {Law}}, \bibinfo {author} {\bibfnamefont
  {X.}~\bibnamefont {Li}}, \bibinfo {author} {\bibfnamefont {F.}~\bibnamefont
  {Foglia}}, \bibinfo {author} {\bibfnamefont {A.}~\bibnamefont {Walsh}},
  \bibinfo {author} {\bibfnamefont {O.~B.}\ \bibnamefont {C}}, \bibinfo
  {author} {\bibfnamefont {J.}~\bibnamefont {Nelson}}, \bibinfo {author}
  {\bibfnamefont {J.~T.}\ \bibnamefont {Cabral}}, \ and\ \bibinfo {author}
  {\bibfnamefont {P.~R.}\ \bibnamefont {Barnes}},\ }\href@noop {} {\bibfield
  {journal} {\bibinfo  {journal} {Nature Communications}\ }\textbf {\bibinfo
  {volume} {6}} (\bibinfo {year} {2015})}\BibitemShut {NoStop}%
\bibitem [{\citenamefont {Even}\ \emph {et~al.}(2016)\citenamefont {Even},
  \citenamefont {Carignano},\ and\ \citenamefont {Katan}}]{Even2016}%
  \BibitemOpen
  \bibfield  {author} {\bibinfo {author} {\bibfnamefont {J.}~\bibnamefont
  {Even}}, \bibinfo {author} {\bibfnamefont {M.}~\bibnamefont {Carignano}}, \
  and\ \bibinfo {author} {\bibfnamefont {C.}~\bibnamefont {Katan}},\
  }\href@noop {} {\bibfield  {journal} {\bibinfo  {journal} {Nanoscale}\
  }\textbf {\bibinfo {volume} {8}} (\bibinfo {year} {2016})}\BibitemShut
  {NoStop}%
\bibitem [{\citenamefont {Brivio}\ \emph {et~al.}(2015)\citenamefont {Brivio},
  \citenamefont {Frost}, \citenamefont {Skelton}, \citenamefont {Jackson},
  \citenamefont {Weber}, \citenamefont {Weller}, \citenamefont {Go\~ni},
  \citenamefont {Leguy}, \citenamefont {Barnes},\ and\ \citenamefont
  {Walsh}}]{Brivio2015}%
  \BibitemOpen
  \bibfield  {author} {\bibinfo {author} {\bibfnamefont {F.}~\bibnamefont
  {Brivio}}, \bibinfo {author} {\bibfnamefont {J.~M.}\ \bibnamefont {Frost}},
  \bibinfo {author} {\bibfnamefont {J.~M.}\ \bibnamefont {Skelton}}, \bibinfo
  {author} {\bibfnamefont {A.~J.}\ \bibnamefont {Jackson}}, \bibinfo {author}
  {\bibfnamefont {O.~J.}\ \bibnamefont {Weber}}, \bibinfo {author}
  {\bibfnamefont {M.~T.}\ \bibnamefont {Weller}}, \bibinfo {author}
  {\bibfnamefont {A.~R.}\ \bibnamefont {Go\~ni}}, \bibinfo {author}
  {\bibfnamefont {A.~M.~A.}\ \bibnamefont {Leguy}}, \bibinfo {author}
  {\bibfnamefont {P.~R.~F.}\ \bibnamefont {Barnes}}, \ and\ \bibinfo {author}
  {\bibfnamefont {A.}~\bibnamefont {Walsh}},\ }\href@noop {} {\bibfield
  {journal} {\bibinfo  {journal} {Phys. Rev. B}\ }\textbf {\bibinfo {volume}
  {92}},\ \bibinfo {pages} {144308} (\bibinfo {year} {2015})}\BibitemShut
  {NoStop}%
\bibitem [{\citenamefont {Zhu}\ \emph {et~al.}(2016)\citenamefont {Zhu},
  \citenamefont {Miyata}, \citenamefont {Fu}, \citenamefont {Wang},
  \citenamefont {Joshi}, \citenamefont {Niesner}, \citenamefont {Williams},
  \citenamefont {Jin},\ and\ \citenamefont {Zhu}}]{Zhu2016}%
  \BibitemOpen
  \bibfield  {author} {\bibinfo {author} {\bibfnamefont {H.}~\bibnamefont
  {Zhu}}, \bibinfo {author} {\bibfnamefont {K.}~\bibnamefont {Miyata}},
  \bibinfo {author} {\bibfnamefont {Y.}~\bibnamefont {Fu}}, \bibinfo {author}
  {\bibfnamefont {J.}~\bibnamefont {Wang}}, \bibinfo {author} {\bibfnamefont
  {P.~P.}\ \bibnamefont {Joshi}}, \bibinfo {author} {\bibfnamefont
  {D.}~\bibnamefont {Niesner}}, \bibinfo {author} {\bibfnamefont {K.~K.}\
  \bibnamefont {Williams}}, \bibinfo {author} {\bibfnamefont {S.}~\bibnamefont
  {Jin}}, \ and\ \bibinfo {author} {\bibfnamefont {X.-Y.}\ \bibnamefont
  {Zhu}},\ }\href@noop {} {\bibfield  {journal} {\bibinfo  {journal} {Science}\
  }\textbf {\bibinfo {volume} {353}},\ \bibinfo {pages} {1409} (\bibinfo {year}
  {2016})}\BibitemShut {NoStop}%
\bibitem [{\citenamefont {T.}(1993)}]{Dove1993}%
  \BibitemOpen
  \bibfield  {author} {\bibinfo {author} {\bibfnamefont {D.~M.}\ \bibnamefont
  {T.}},\ }\href@noop {} {\emph {\bibinfo {title} {Introduction to Lattice
  Dynamics}}}\ (\bibinfo  {publisher} {Cambridge University Press, Cambridge,
  UK},\ \bibinfo {year} {1993})\BibitemShut {NoStop}%
\bibitem [{\citenamefont {Shirane}\ \emph {et~al.}(2004)\citenamefont
  {Shirane}, \citenamefont {Shapiro},\ and\ \citenamefont
  {Tranquada}}]{Shirane2004}%
  \BibitemOpen
  \bibfield  {author} {\bibinfo {author} {\bibfnamefont {G.}~\bibnamefont
  {Shirane}}, \bibinfo {author} {\bibfnamefont {S.~M.}\ \bibnamefont
  {Shapiro}}, \ and\ \bibinfo {author} {\bibfnamefont {J.~M.}\ \bibnamefont
  {Tranquada}},\ }\href@noop {} {\emph {\bibinfo {title} {Neutron Scattering
  with a Triple-Axis Spectrometer}}}\ (\bibinfo  {publisher} {Cambridge
  University Press, Cambridge, UK},\ \bibinfo {year} {2004})\BibitemShut
  {NoStop}%
\bibitem [{\citenamefont {Bee}(1988)}]{Bee2004}%
  \BibitemOpen
  \bibfield  {author} {\bibinfo {author} {\bibfnamefont {M.}~\bibnamefont
  {Bee}},\ }\href@noop {} {\emph {\bibinfo {title} {Quasielastic Neutron
  Scattering}}}\ (\bibinfo  {publisher} {Adam Hilger, Bristol, UK},\ \bibinfo
  {year} {1988})\BibitemShut {NoStop}%
\bibitem [{\citenamefont {L\'etoublon}\ \emph {et~al.}(2016)\citenamefont
  {L\'etoublon}, \citenamefont {Paofai}, \citenamefont {Ruffl\'e},
  \citenamefont {Bourges}, \citenamefont {Hehlen}, \citenamefont {Michel},
  \citenamefont {Ecolivet}, \citenamefont {Durand}, \citenamefont {Cordier},
  \citenamefont {Katan},\ and\ \citenamefont {Even}}]{Letoublon2016}%
  \BibitemOpen
  \bibfield  {author} {\bibinfo {author} {\bibfnamefont {A.}~\bibnamefont
  {L\'etoublon}}, \bibinfo {author} {\bibfnamefont {S.}~\bibnamefont {Paofai}},
  \bibinfo {author} {\bibfnamefont {B.}~\bibnamefont {Ruffl\'e}}, \bibinfo
  {author} {\bibfnamefont {P.}~\bibnamefont {Bourges}}, \bibinfo {author}
  {\bibfnamefont {B.}~\bibnamefont {Hehlen}}, \bibinfo {author} {\bibfnamefont
  {T.}~\bibnamefont {Michel}}, \bibinfo {author} {\bibfnamefont
  {C.}~\bibnamefont {Ecolivet}}, \bibinfo {author} {\bibfnamefont
  {O.}~\bibnamefont {Durand}}, \bibinfo {author} {\bibfnamefont
  {S.}~\bibnamefont {Cordier}}, \bibinfo {author} {\bibfnamefont
  {C.}~\bibnamefont {Katan}}, \ and\ \bibinfo {author} {\bibfnamefont
  {J.}~\bibnamefont {Even}},\ }\href@noop {} {\bibfield  {journal} {\bibinfo
  {journal} {J. Phys. Chem. Lett.}\ }\textbf {\bibinfo {volume} {7}},\ \bibinfo
  {pages} {3776} (\bibinfo {year} {2016})}\BibitemShut {NoStop}%
\bibitem [{\citenamefont {Ferreira}\ \emph {et~al.}(2018)\citenamefont
  {Ferreira}, \citenamefont {L\'etoublon}, \citenamefont {Paofai},
  \citenamefont {Raymond}, \citenamefont {Ecolivet}, \citenamefont {Ruffl\'e},
  \citenamefont {Cordier}, \citenamefont {Katan}, \citenamefont {Saidaminov},
  \citenamefont {Zhumekenov}, \citenamefont {Bakr}, \citenamefont {Even},\ and\
  \citenamefont {Bourges}}]{Ferreira2018}%
  \BibitemOpen
  \bibfield  {author} {\bibinfo {author} {\bibfnamefont {A.~C.}\ \bibnamefont
  {Ferreira}}, \bibinfo {author} {\bibfnamefont {A.}~\bibnamefont
  {L\'etoublon}}, \bibinfo {author} {\bibfnamefont {S.}~\bibnamefont {Paofai}},
  \bibinfo {author} {\bibfnamefont {S.}~\bibnamefont {Raymond}}, \bibinfo
  {author} {\bibfnamefont {C.}~\bibnamefont {Ecolivet}}, \bibinfo {author}
  {\bibfnamefont {B.}~\bibnamefont {Ruffl\'e}}, \bibinfo {author}
  {\bibfnamefont {S.}~\bibnamefont {Cordier}}, \bibinfo {author} {\bibfnamefont
  {C.}~\bibnamefont {Katan}}, \bibinfo {author} {\bibfnamefont {M.~I.}\
  \bibnamefont {Saidaminov}}, \bibinfo {author} {\bibfnamefont {A.~A.}\
  \bibnamefont {Zhumekenov}}, \bibinfo {author} {\bibfnamefont {O.~M.}\
  \bibnamefont {Bakr}}, \bibinfo {author} {\bibfnamefont {J.}~\bibnamefont
  {Even}}, \ and\ \bibinfo {author} {\bibfnamefont {P.}~\bibnamefont
  {Bourges}},\ }\href@noop {} {\bibfield  {journal} {\bibinfo  {journal} {Phys.
  Rev. Lett.}\ }\textbf {\bibinfo {volume} {121}},\ \bibinfo {pages} {085502}
  (\bibinfo {year} {2018})}\BibitemShut {NoStop}%
\bibitem [{\citenamefont {Beecher}\ \emph {et~al.}(2016)\citenamefont
  {Beecher}, \citenamefont {Semonin}, \citenamefont {Skelton}, \citenamefont
  {Frost}, \citenamefont {Terban}, \citenamefont {Zhai}, \citenamefont
  {Alatas}, \citenamefont {Owen}, \citenamefont {Walsh},\ and\ \citenamefont
  {Billinge}}]{Beecher2016}%
  \BibitemOpen
  \bibfield  {author} {\bibinfo {author} {\bibfnamefont {A.~N.}\ \bibnamefont
  {Beecher}}, \bibinfo {author} {\bibfnamefont {O.~E.}\ \bibnamefont
  {Semonin}}, \bibinfo {author} {\bibfnamefont {J.~M.}\ \bibnamefont
  {Skelton}}, \bibinfo {author} {\bibfnamefont {J.~M.}\ \bibnamefont {Frost}},
  \bibinfo {author} {\bibfnamefont {M.~W.}\ \bibnamefont {Terban}}, \bibinfo
  {author} {\bibfnamefont {H.}~\bibnamefont {Zhai}}, \bibinfo {author}
  {\bibfnamefont {A.}~\bibnamefont {Alatas}}, \bibinfo {author} {\bibfnamefont
  {J.~S.}\ \bibnamefont {Owen}}, \bibinfo {author} {\bibfnamefont
  {A.}~\bibnamefont {Walsh}}, \ and\ \bibinfo {author} {\bibfnamefont
  {S.~J.~L.}\ \bibnamefont {Billinge}},\ }\href@noop {} {\bibfield  {journal}
  {\bibinfo  {journal} {ACS Energy Lett.}\ }\textbf {\bibinfo {volume} {1}}
  (\bibinfo {year} {2016})}\BibitemShut {NoStop}%
\bibitem [{\citenamefont {Weber}\ \emph {et~al.}(2017)\citenamefont {Weber},
  \citenamefont {Roessli}, \citenamefont {Stock}, \citenamefont {Keller},
  \citenamefont {Schmalzl}, \citenamefont {Bourdarot}, \citenamefont {Georgii},
  \citenamefont {Ewings}, \citenamefont {Perry},\ and\ \citenamefont
  {B\"oni}}]{Weber2017}%
  \BibitemOpen
  \bibfield  {author} {\bibinfo {author} {\bibfnamefont {T.}~\bibnamefont
  {Weber}}, \bibinfo {author} {\bibfnamefont {B.}~\bibnamefont {Roessli}},
  \bibinfo {author} {\bibfnamefont {C.}~\bibnamefont {Stock}}, \bibinfo
  {author} {\bibfnamefont {T.}~\bibnamefont {Keller}}, \bibinfo {author}
  {\bibfnamefont {K.}~\bibnamefont {Schmalzl}}, \bibinfo {author}
  {\bibfnamefont {F.}~\bibnamefont {Bourdarot}}, \bibinfo {author}
  {\bibfnamefont {R.}~\bibnamefont {Georgii}}, \bibinfo {author} {\bibfnamefont
  {R.~A.}\ \bibnamefont {Ewings}}, \bibinfo {author} {\bibfnamefont {R.~S.}\
  \bibnamefont {Perry}}, \ and\ \bibinfo {author} {\bibfnamefont
  {P.}~\bibnamefont {B\"oni}},\ }\href@noop {} {\bibfield  {journal} {\bibinfo
  {journal} {Phys. Rev. B}\ }\textbf {\bibinfo {volume} {96}},\ \bibinfo
  {pages} {184301} (\bibinfo {year} {2017})}\BibitemShut {NoStop}%
\bibitem [{\citenamefont {Collins}(1989)}]{Collins1989}%
  \BibitemOpen
  \bibfield  {author} {\bibinfo {author} {\bibfnamefont {M.}~\bibnamefont
  {Collins}},\ }\href@noop {} {\emph {\bibinfo {title} {Magnetic critical
  scattering}}}\ (\bibinfo  {publisher} {Oxford University Press},\ \bibinfo
  {year} {1989})\BibitemShut {NoStop}%
\bibitem [{\citenamefont {Shirane}\ \emph {et~al.}(1967)\citenamefont
  {Shirane}, \citenamefont {Fazer}, \citenamefont {Minkiewicz}, \citenamefont
  {Leake},\ and\ \citenamefont {Linz}}]{Shirane1967}%
  \BibitemOpen
  \bibfield  {author} {\bibinfo {author} {\bibfnamefont {G.}~\bibnamefont
  {Shirane}}, \bibinfo {author} {\bibfnamefont {B.~C.}\ \bibnamefont {Fazer}},
  \bibinfo {author} {\bibfnamefont {V.~J.}\ \bibnamefont {Minkiewicz}},
  \bibinfo {author} {\bibfnamefont {J.~A.}\ \bibnamefont {Leake}}, \ and\
  \bibinfo {author} {\bibfnamefont {A.}~\bibnamefont {Linz}},\ }\href@noop {}
  {\bibfield  {journal} {\bibinfo  {journal} {Phys. Rev. Lett.}\ }\textbf
  {\bibinfo {volume} {19}},\ \bibinfo {pages} {234} (\bibinfo {year}
  {1967})}\BibitemShut {NoStop}%
\bibitem [{\citenamefont {Lynn}\ \emph {et~al.}(1978)\citenamefont {Lynn},
  \citenamefont {Patterson}, \citenamefont {Shirane},\ and\ \citenamefont
  {Wheeler}}]{Lynn1978}%
  \BibitemOpen
  \bibfield  {author} {\bibinfo {author} {\bibfnamefont {J.~W.}\ \bibnamefont
  {Lynn}}, \bibinfo {author} {\bibfnamefont {H.~H.}\ \bibnamefont {Patterson}},
  \bibinfo {author} {\bibfnamefont {G.}~\bibnamefont {Shirane}}, \ and\
  \bibinfo {author} {\bibfnamefont {R.~G.}\ \bibnamefont {Wheeler}},\
  }\href@noop {} {\bibfield  {journal} {\bibinfo  {journal} {Solid State
  Commun.}\ }\textbf {\bibinfo {volume} {27}},\ \bibinfo {pages} {859}
  (\bibinfo {year} {1978})}\BibitemShut {NoStop}%
\bibitem [{\citenamefont {Lee}\ \emph {et~al.}(2016)\citenamefont {Lee},
  \citenamefont {Bristowe}, \citenamefont {Lee}, \citenamefont {Lee},
  \citenamefont {Bristowe}, \citenamefont {Cheetham},\ and\ \citenamefont
  {Jang}}]{Lee2016}%
  \BibitemOpen
  \bibfield  {author} {\bibinfo {author} {\bibfnamefont {J.~H.}\ \bibnamefont
  {Lee}}, \bibinfo {author} {\bibfnamefont {N.~C.}\ \bibnamefont {Bristowe}},
  \bibinfo {author} {\bibfnamefont {J.~H.}\ \bibnamefont {Lee}}, \bibinfo
  {author} {\bibfnamefont {S.~H.}\ \bibnamefont {Lee}}, \bibinfo {author}
  {\bibfnamefont {P.~D.}\ \bibnamefont {Bristowe}}, \bibinfo {author}
  {\bibfnamefont {A.~K.}\ \bibnamefont {Cheetham}}, \ and\ \bibinfo {author}
  {\bibfnamefont {H.~M.}\ \bibnamefont {Jang}},\ }\href@noop {} {\bibfield
  {journal} {\bibinfo  {journal} {Chem. Matter.}\ }\textbf {\bibinfo {volume}
  {28}},\ \bibinfo {pages} {4259} (\bibinfo {year} {2016})}\BibitemShut
  {NoStop}%
\bibitem [{\citenamefont {Swainson}\ \emph {et~al.}(2003)\citenamefont
  {Swainson}, \citenamefont {Hammond}, \citenamefont {Soulli{\`e}re},
  \citenamefont {Knop},\ and\ \citenamefont {Massa}}]{Swainson2003}%
  \BibitemOpen
  \bibfield  {author} {\bibinfo {author} {\bibfnamefont {I.~P.}\ \bibnamefont
  {Swainson}}, \bibinfo {author} {\bibfnamefont {R.~P.}\ \bibnamefont
  {Hammond}}, \bibinfo {author} {\bibfnamefont {C.}~\bibnamefont
  {Soulli{\`e}re}}, \bibinfo {author} {\bibfnamefont {O.}~\bibnamefont {Knop}},
  \ and\ \bibinfo {author} {\bibfnamefont {W.}~\bibnamefont {Massa}},\
  }\href@noop {} {\bibfield  {journal} {\bibinfo  {journal} {J. Solid State
  Chem.}\ }\textbf {\bibinfo {volume} {176}},\ \bibinfo {pages} {97} (\bibinfo
  {year} {2003})}\BibitemShut {NoStop}%
\bibitem [{\citenamefont {Shirane}\ and\ \citenamefont
  {Yamada}(1969)}]{Shirane1969}%
  \BibitemOpen
  \bibfield  {author} {\bibinfo {author} {\bibfnamefont {G.}~\bibnamefont
  {Shirane}}\ and\ \bibinfo {author} {\bibfnamefont {Y.}~\bibnamefont
  {Yamada}},\ }\href@noop {} {\bibfield  {journal} {\bibinfo  {journal} {Phys.
  Rev.}\ }\textbf {\bibinfo {volume} {177}},\ \bibinfo {pages} {858} (\bibinfo
  {year} {1969})}\BibitemShut {NoStop}%
\bibitem [{\citenamefont {Shapiro}\ \emph {et~al.}(1972)\citenamefont
  {Shapiro}, \citenamefont {Axe}, \citenamefont {Shirane},\ and\ \citenamefont
  {Riste}}]{Shapiro1972}%
  \BibitemOpen
  \bibfield  {author} {\bibinfo {author} {\bibfnamefont {S.~M.}\ \bibnamefont
  {Shapiro}}, \bibinfo {author} {\bibfnamefont {J.}~\bibnamefont {Axe}},
  \bibinfo {author} {\bibfnamefont {G.}~\bibnamefont {Shirane}}, \ and\
  \bibinfo {author} {\bibfnamefont {T.}~\bibnamefont {Riste}},\ }\href@noop {}
  {\bibfield  {journal} {\bibinfo  {journal} {Phys. Rev. B}\ }\textbf {\bibinfo
  {volume} {6}},\ \bibinfo {pages} {4332} (\bibinfo {year} {1972})}\BibitemShut
  {NoStop}%
\bibitem [{\citenamefont {Hagen}\ \emph {et~al.}(1992)\citenamefont {Hagen},
  \citenamefont {Dove}, \citenamefont {Harris}, \citenamefont {Steigenberger},\
  and\ \citenamefont {Powell}}]{Hagen1992}%
  \BibitemOpen
  \bibfield  {author} {\bibinfo {author} {\bibfnamefont {M.}~\bibnamefont
  {Hagen}}, \bibinfo {author} {\bibfnamefont {M.~T.}\ \bibnamefont {Dove}},
  \bibinfo {author} {\bibfnamefont {M.~J.}\ \bibnamefont {Harris}}, \bibinfo
  {author} {\bibfnamefont {U.}~\bibnamefont {Steigenberger}}, \ and\ \bibinfo
  {author} {\bibfnamefont {B.~M.}\ \bibnamefont {Powell}},\ }\href@noop {}
  {\bibfield  {journal} {\bibinfo  {journal} {Physica B}\ }\textbf {\bibinfo
  {volume} {180 \& 181}},\ \bibinfo {pages} {276} (\bibinfo {year}
  {1992})}\BibitemShut {NoStop}%
\bibitem [{\citenamefont {Cowley}(1980)}]{Cowley1980}%
  \BibitemOpen
  \bibfield  {author} {\bibinfo {author} {\bibfnamefont {R.~A.}\ \bibnamefont
  {Cowley}},\ }\href@noop {} {\bibfield  {journal} {\bibinfo  {journal}
  {Advances in physics}\ }\textbf {\bibinfo {volume} {29}},\ \bibinfo {pages}
  {1} (\bibinfo {year} {1980})}\BibitemShut {NoStop}%
\bibitem [{\citenamefont {Cowley}(1976)}]{Cowley1976}%
  \BibitemOpen
  \bibfield  {author} {\bibinfo {author} {\bibfnamefont {R.~A.}\ \bibnamefont
  {Cowley}},\ }\href@noop {} {\bibfield  {journal} {\bibinfo  {journal} {Phys.
  Rev. B}\ }\textbf {\bibinfo {volume} {13}},\ \bibinfo {pages} {4877}
  (\bibinfo {year} {1976})}\BibitemShut {NoStop}%
\bibitem [{\citenamefont {Bechtel}\ and\ \citenamefont {Van~der
  Ven}(2018)}]{Bechtel2018}%
  \BibitemOpen
  \bibfield  {author} {\bibinfo {author} {\bibfnamefont {J.~S.}\ \bibnamefont
  {Bechtel}}\ and\ \bibinfo {author} {\bibfnamefont {A.}~\bibnamefont {Van~der
  Ven}},\ }\href@noop {} {\bibfield  {journal} {\bibinfo  {journal} {Phys. Rev.
  Mater.}\ }\textbf {\bibinfo {volume} {2}} (\bibinfo {year}
  {2018})}\BibitemShut {NoStop}%
\bibitem [{\citenamefont {Shirane}(1974)}]{Shirane1974}%
  \BibitemOpen
  \bibfield  {author} {\bibinfo {author} {\bibfnamefont {G.}~\bibnamefont
  {Shirane}},\ }\href@noop {} {\bibfield  {journal} {\bibinfo  {journal} {Rev.
  Mod. Phys.}\ }\textbf {\bibinfo {volume} {46}},\ \bibinfo {pages} {437}
  (\bibinfo {year} {1974})}\BibitemShut {NoStop}%
\bibitem [{\citenamefont {Scott}(1974)}]{Scott1974}%
  \BibitemOpen
  \bibfield  {author} {\bibinfo {author} {\bibfnamefont {J.~F.}\ \bibnamefont
  {Scott}},\ }\href@noop {} {\bibfield  {journal} {\bibinfo  {journal} {Rev.
  Mod. Phys.}\ }\textbf {\bibinfo {volume} {46}},\ \bibinfo {pages} {83}
  (\bibinfo {year} {1974})}\BibitemShut {NoStop}%
\bibitem [{\citenamefont {Shirane}\ \emph {et~al.}(1970)\citenamefont
  {Shirane}, \citenamefont {Axe}, \citenamefont {Harada},\ and\ \citenamefont
  {Remeika}}]{Shirane1970}%
  \BibitemOpen
  \bibfield  {author} {\bibinfo {author} {\bibfnamefont {G.}~\bibnamefont
  {Shirane}}, \bibinfo {author} {\bibfnamefont {J.~D.}\ \bibnamefont {Axe}},
  \bibinfo {author} {\bibfnamefont {J.}~\bibnamefont {Harada}}, \ and\ \bibinfo
  {author} {\bibfnamefont {J.~P.}\ \bibnamefont {Remeika}},\ }\href@noop {}
  {\bibfield  {journal} {\bibinfo  {journal} {Phys. Rev. B}\ }\textbf {\bibinfo
  {volume} {2}},\ \bibinfo {pages} {155} (\bibinfo {year} {1970})}\BibitemShut
  {NoStop}%
\bibitem [{\citenamefont {Kempa}\ \emph {et~al.}(2006)\citenamefont {Kempa},
  \citenamefont {Hlinka}, \citenamefont {Kulda}, \citenamefont {Bourges},
  \citenamefont {Kania},\ and\ \citenamefont {Petzelt}}]{Kempa2006}%
  \BibitemOpen
  \bibfield  {author} {\bibinfo {author} {\bibfnamefont {M.}~\bibnamefont
  {Kempa}}, \bibinfo {author} {\bibfnamefont {J.}~\bibnamefont {Hlinka}},
  \bibinfo {author} {\bibfnamefont {J.}~\bibnamefont {Kulda}}, \bibinfo
  {author} {\bibfnamefont {P.}~\bibnamefont {Bourges}}, \bibinfo {author}
  {\bibfnamefont {A.}~\bibnamefont {Kania}}, \ and\ \bibinfo {author}
  {\bibfnamefont {J.}~\bibnamefont {Petzelt}},\ }\href@noop {} {\bibfield
  {journal} {\bibinfo  {journal} {Phase Transisions}\ }\textbf {\bibinfo
  {volume} {79}},\ \bibinfo {pages} {351} (\bibinfo {year} {2006})}\BibitemShut
  {NoStop}%
\bibitem [{\citenamefont {Yamada}\ \emph
  {et~al.}(1974{\natexlab{a}})\citenamefont {Yamada}, \citenamefont {Noda},
  \citenamefont {Axe},\ and\ \citenamefont {Shirane}}]{Yamada1974}%
  \BibitemOpen
  \bibfield  {author} {\bibinfo {author} {\bibfnamefont {Y.}~\bibnamefont
  {Yamada}}, \bibinfo {author} {\bibfnamefont {Y.}~\bibnamefont {Noda}},
  \bibinfo {author} {\bibfnamefont {J.~D.}\ \bibnamefont {Axe}}, \ and\
  \bibinfo {author} {\bibfnamefont {G.}~\bibnamefont {Shirane}},\ }\href@noop
  {} {\bibfield  {journal} {\bibinfo  {journal} {Phys. Rev. B}\ }\textbf
  {\bibinfo {volume} {9}},\ \bibinfo {pages} {4429} (\bibinfo {year}
  {1974}{\natexlab{a}})}\BibitemShut {NoStop}%
\bibitem [{\citenamefont {Yamada}\ \emph {et~al.}(1972)\citenamefont {Yamada},
  \citenamefont {Mori},\ and\ \citenamefont {Noda}}]{Yamada1972}%
  \BibitemOpen
  \bibfield  {author} {\bibinfo {author} {\bibfnamefont {Y.}~\bibnamefont
  {Yamada}}, \bibinfo {author} {\bibfnamefont {M.}~\bibnamefont {Mori}}, \ and\
  \bibinfo {author} {\bibfnamefont {Y.}~\bibnamefont {Noda}},\ }\href@noop {}
  {\bibfield  {journal} {\bibinfo  {journal} {J. Phys. Soc. Jpn}\ }\textbf
  {\bibinfo {volume} {32}},\ \bibinfo {pages} {1565} (\bibinfo {year}
  {1972})}\BibitemShut {NoStop}%
\bibitem [{\citenamefont {Halperin}\ and\ \citenamefont
  {Varma}(1976)}]{Halperin1976}%
  \BibitemOpen
  \bibfield  {author} {\bibinfo {author} {\bibfnamefont {B.~I.}\ \bibnamefont
  {Halperin}}\ and\ \bibinfo {author} {\bibfnamefont {C.~M.}\ \bibnamefont
  {Varma}},\ }\href@noop {} {\bibfield  {journal} {\bibinfo  {journal} {Phys.
  Rev. B}\ }\textbf {\bibinfo {volume} {14}},\ \bibinfo {pages} {4030}
  (\bibinfo {year} {1976})}\BibitemShut {NoStop}%
\bibitem [{\citenamefont {Aubry}(1975)}]{Aubry1975}%
  \BibitemOpen
  \bibfield  {author} {\bibinfo {author} {\bibfnamefont {S.}~\bibnamefont
  {Aubry}},\ }\href@noop {} {\bibfield  {journal} {\bibinfo  {journal} {J.
  Chem. Phys.}\ }\textbf {\bibinfo {volume} {62}},\ \bibinfo {pages} {3217}
  (\bibinfo {year} {1975})}\BibitemShut {NoStop}%
\bibitem [{\citenamefont {Bruce}(1980)}]{Bruce1980bis}%
  \BibitemOpen
  \bibfield  {author} {\bibinfo {author} {\bibfnamefont {A.~D.}\ \bibnamefont
  {Bruce}},\ }\href@noop {} {\bibfield  {journal} {\bibinfo  {journal} {Adv.
  Phys.}\ }\textbf {\bibinfo {volume} {29}},\ \bibinfo {pages} {111} (\bibinfo
  {year} {1980})}\BibitemShut {NoStop}%
\bibitem [{\citenamefont {Bruce}\ and\ \citenamefont
  {Cowley}(1980)}]{Bruce1980}%
  \BibitemOpen
  \bibfield  {author} {\bibinfo {author} {\bibfnamefont {A.~D.}\ \bibnamefont
  {Bruce}}\ and\ \bibinfo {author} {\bibfnamefont {R.~A.}\ \bibnamefont
  {Cowley}},\ }\href@noop {} {\bibfield  {journal} {\bibinfo  {journal} {Adv.
  Phys.}\ }\textbf {\bibinfo {volume} {29}},\ \bibinfo {pages} {219} (\bibinfo
  {year} {1980})}\BibitemShut {NoStop}%
\bibitem [{\citenamefont {Yamada}\ \emph
  {et~al.}(1974{\natexlab{b}})\citenamefont {Yamada}, \citenamefont
  {Takatera},\ and\ \citenamefont {Huber}}]{Huber1974}%
  \BibitemOpen
  \bibfield  {author} {\bibinfo {author} {\bibfnamefont {Y.}~\bibnamefont
  {Yamada}}, \bibinfo {author} {\bibfnamefont {H.}~\bibnamefont {Takatera}}, \
  and\ \bibinfo {author} {\bibfnamefont {D.~L.}\ \bibnamefont {Huber}},\
  }\href@noop {} {\bibfield  {journal} {\bibinfo  {journal} {J. Phys. Soc.
  Jpn}\ }\textbf {\bibinfo {volume} {36}},\ \bibinfo {pages} {641} (\bibinfo
  {year} {1974}{\natexlab{b}})}\BibitemShut {NoStop}%
\bibitem [{\citenamefont {Comin}\ \emph {et~al.}(2016)\citenamefont {Comin},
  \citenamefont {Crawford}, \citenamefont {Said}, \citenamefont {Herron},
  \citenamefont {Guise}, \citenamefont {Wang}, \citenamefont {Whitfield},
  \citenamefont {Jain}, \citenamefont {Gong}, \citenamefont {McGaughey},\ and\
  \citenamefont {Sargent}}]{Comin2016}%
  \BibitemOpen
  \bibfield  {author} {\bibinfo {author} {\bibfnamefont {R.}~\bibnamefont
  {Comin}}, \bibinfo {author} {\bibfnamefont {M.~K.}\ \bibnamefont {Crawford}},
  \bibinfo {author} {\bibfnamefont {A.~H.}\ \bibnamefont {Said}}, \bibinfo
  {author} {\bibfnamefont {N.}~\bibnamefont {Herron}}, \bibinfo {author}
  {\bibfnamefont {W.~E.}\ \bibnamefont {Guise}}, \bibinfo {author}
  {\bibfnamefont {X.}~\bibnamefont {Wang}}, \bibinfo {author} {\bibfnamefont
  {P.~S.}\ \bibnamefont {Whitfield}}, \bibinfo {author} {\bibfnamefont
  {A.}~\bibnamefont {Jain}}, \bibinfo {author} {\bibfnamefont {X.}~\bibnamefont
  {Gong}}, \bibinfo {author} {\bibfnamefont {A.~J.~H.}\ \bibnamefont
  {McGaughey}}, \ and\ \bibinfo {author} {\bibfnamefont {E.~H.}\ \bibnamefont
  {Sargent}},\ }\href@noop {} {\bibfield  {journal} {\bibinfo  {journal} {Phys.
  Rev. B}\ }\textbf {\bibinfo {volume} {94}},\ \bibinfo {pages} {094301}
  (\bibinfo {year} {2016})}\BibitemShut {NoStop}%
\bibitem [{\citenamefont {Stock}\ \emph {et~al.}(2010)\citenamefont {Stock},
  \citenamefont {Van~Eijck}, \citenamefont {Fouquet}, \citenamefont
  {Maccarini}, \citenamefont {Gehring}, \citenamefont {Xu}, \citenamefont
  {Luo}, \citenamefont {Zhao}, \citenamefont {Li},\ and\ \citenamefont
  {Viehland}}]{Stock2010}%
  \BibitemOpen
  \bibfield  {author} {\bibinfo {author} {\bibfnamefont {C.}~\bibnamefont
  {Stock}}, \bibinfo {author} {\bibfnamefont {L.}~\bibnamefont {Van~Eijck}},
  \bibinfo {author} {\bibfnamefont {P.}~\bibnamefont {Fouquet}}, \bibinfo
  {author} {\bibfnamefont {M.}~\bibnamefont {Maccarini}}, \bibinfo {author}
  {\bibfnamefont {P.~M.}\ \bibnamefont {Gehring}}, \bibinfo {author}
  {\bibfnamefont {G.}~\bibnamefont {Xu}}, \bibinfo {author} {\bibfnamefont
  {H.}~\bibnamefont {Luo}}, \bibinfo {author} {\bibfnamefont {X.}~\bibnamefont
  {Zhao}}, \bibinfo {author} {\bibfnamefont {J.-F.}\ \bibnamefont {Li}}, \ and\
  \bibinfo {author} {\bibfnamefont {D.}~\bibnamefont {Viehland}},\ }\href@noop
  {} {\bibfield  {journal} {\bibinfo  {journal} {Phys. Rev. B}\ }\textbf
  {\bibinfo {volume} {81}},\ \bibinfo {pages} {144127} (\bibinfo {year}
  {2010})}\BibitemShut {NoStop}%
\end{thebibliography}%

\end{document}